\documentclass[12pt]{article}
\usepackage{amsmath}
\usepackage{graphicx}
\usepackage{enumerate}
\usepackage{natbib}
\usepackage{url} % not crucial - just used below for the URL
\usepackage{booktabs}
\usepackage{graphicx}
\usepackage{caption,subcaption}
\usepackage{amssymb,amsfonts,amsthm}
\usepackage{tikz}
\usetikzlibrary{positioning}
\usetikzlibrary{arrows,arrows.meta}
\RequirePackage[colorlinks,citecolor=blue,urlcolor=blue]{hyperref}

\hypersetup{backref, final=false, pdfpagemode=FullScreen,  colorlinks=true}

\usepackage{amssymb}
\usepackage{longtable}
\usepackage{lineno,hyperref}
\usepackage[section]{placeins}
\usepackage{multirow}
\usepackage{array}
\makeatletter
\def\singlespace{\def\baselinestretch{0.7}\@normalsize}
\def\beqn{\begin{eqnarray}}
	\def\eeqn{\end{eqnarray}}
\def\beqns{\begin{eqnarray*}}
	\def\eeqns{\end{eqnarray*}}

\newtheorem{Theorem}{Theorem}%[section]
\newtheorem{Remark}{Remark}%[section]
\newtheorem{Example}{Example}%[section]
\newtheorem{Proposition}{Proposition}%[section]

\newcounter{thm}

\newcommand{\bldsi}[1]{\mbox{\tiny  $#1$}}
\newcommand{\bldsr}[1]{\mbox{\footnotesize  $#1$}}

\def\bOmega{\mbox{\boldmath{$\Omega$}}}

\def\bPsi{\mbox{\boldmath{$\Psi$}}}

\def\bSigma{\mbox{\boldmath{$\Sigma$}}}

\def\bbeta{\mbox{\boldmath{$\beta$}}}

\def\bepsilon{\mbox{\boldmath{$\varepsilon$}}}

\def\bX{\mathbf X}
\def\bY{\mathbf Y}

\def\bA{\mathbf A}
\def\bB{\mathbf B}

\def\bD{\mathbf D}
\def\bE{\mathbf E}

\def\bI{\mathbf I}

\def\bP{\mathbf P}

\def\bU{\mathbf U}
\def\bV{\mathbf V}

\def\ba{\mathbf a}
\def\bb{\mathbf b}

\def\mf{\mathbf f}

\def\bw{\mathbf w}
\def\bx{\mathbf x}

\def\bu{\mathbf u}

\def\b0{\mathbf 0}

\makeatletter
\newcommand\sixteen{\@setfontsize\sixteen{15pt}{6}}
\renewcommand{\maketitle}{\bgroup\setlength{\parindent}{15pt}
	\begin{flushleft}
		\sixteen\bfseries \@title
		\medskip
	\end{flushleft}
	\begin{flushleft}
		\@author
		\medskip
	\end{flushleft}
}
\makeatother
\title{\bf \quad Decorrelated forward regression for high dimensional data \\ \quad analysis}
\author{\hspace{0.7cm}Xuejun Jiang$^1$, Yue Ma$^1$ and Haofeng Wang$^2$  \\ 
	\vspace{0.2cm}
	\hspace{ 0.6cm}\textsl{ $^1$Department of Statistics and Data Science,
		Southern University of Science and} \\
	\hspace{ 0.6cm} \textsl{Technology, China.}\\
	\vspace{0.2cm}
	\hspace{ 0.6cm} 
	\textsl{$^2$Department of Mathematics, Hong Kong Baptist University, Hong Kong}
}

\date{}
\begin{document}
	
	\def\spacingset#1{\renewcommand{\baselinestretch}%
		{#1}\small\normalsize} \spacingset{1}
	
	\maketitle

	\begin{abstract}
		Forward regression is a crucial methodology for automatically identifying important predictors from a large pool of potential covariates. In contexts with moderate predictor correlation, forward selection techniques can achieve screening consistency.  However, this property gradually becomes invalid in the presence of substantially correlated variables, especially in high-dimensional datasets where strong correlations exist among predictors. This dilemma is encountered by other model selection methods in literature as well. To address these challenges, we introduce a novel decorrelated forward (DF) selection framework for generalized mean regression models, including prevalent models, such as linear, logistic, Poisson, and quasi likelihood. The DF selection framework stands out because of its ability to convert generalized mean regression models into linear ones, thus providing a clear interpretation of the forward selection process. It also offers a closed-form expression for forward iteration, to improve practical applicability and efficiency. Theoretically, we establish the screening consistency of DF  selection and determine the upper bound of the selected submodel's size. To reduce computational burden, we  develop a thresholding DF algorithm that provides a stopping rule for the forward-searching process. Simulations and two real data applications show the outstanding performance of our method compared with some existing model selection methods.
		\vspace{0.2cm}
		
	\end{abstract}

	\spacingset{1.75} % DON'T change the spacing!

	\section{Introduction}

 	\subsection{Problem formulation}
Contemporary data-driven scientific research frequently encounters datasets characterized by ultra-high dimensions wherein variables exhibit substantial intercorrelation. The prevalence of ultra-high-dimensional datasets with substantial intercorrelation is evident across various domains, including genetics, econometrics, and experimental physics. Model selection is a pivotal methodology for the analysis of ultra-high-dimensional datasets.
Classical model selection methods, such as least absolute shrinkage and selection operator (LASSO; \citealp{T96}), smoothly clipped absolute deviation (SCAD; \citealp{FL01}), measure-correlate-predict (MCP; \citealp{Z10}), sure independent screening (SIS; \citealp{FL08}), and forward regression (FR; \citealp{W09, CHZ16}), are invalid when confronted with substantially correlated variables. In addition, classical model selection methods rely on specific model assumptions, which may be violated in practical scenarios. To address this limitation, researchers have proposed some model-free variable selection methods based on correlation measures \citep{Zel11,Z20}. To achieve screening consistency,  they placed some restrictions on  the correlations between the response and  features or among predictors, which may be violated when substantially correlated variables are present. This situation highlights the urgent need to develop a generalized framework that can perform feature selection in contexts where predictors exhibit strong correlation.
\vspace{-0.2cm}
\subsection{Main results and contributions} 
Our aim is to develop a novel model selection framework for mean regression models, where $Y \in \mathbb{R}$ represents the response variable and $\mathbf{x} \in \mathbb{R}^{p_n}$ is a vector of covariates. These models are described by the equation:
\begin{equation}\label{b11}
	Y=E(Y|\bx)+ \varepsilon
\end{equation}
with the condition  $g(E(Y|\bx))=\bx^\top\bbeta^*$, where $g(\cdot)$ is a known link function,
$\bbeta^* \in \mathbb{R}^{p_n}$ is an unknown sparse  coefficient vector, and $\varepsilon$ is a random error satisfying
$E(\varepsilon\mid \bx)=0$. 
Model \eqref{b11} represents a comprehensive generalization of the prevalent models frequently referenced in literature:
(i) Canonical exponential generalized linear models (GLMs) incorporate a specified density function given by
	\begin{equation}\label{b1b1}
		f(Y \mid \bx, \bbeta^*, \phi)=\exp \left[\phi^{-1}\left\{Y \bx^{\top} \bbeta^*-b\left(\bx^{\top} \bbeta^*\right)\right\}+c(Y, \phi)\right]. 
	\end{equation}
	GLMs are  special cases of model \eqref{b11} when we select $g=(b')^{-1}$;   
 (ii) Quasi-likelihood models (QLMs) assume the following moment condition:
	\begin{equation}
		E(Y\mid \bx)=\mu(\bx^\top\bbeta^*) \ \ \text{and} \ \ \text{Var}(Y\mid \bx)=
		\phi V(\mu(\bx^\top\bbeta)),
	\end{equation}
	where $\mu(\cdot)$ is a known mean function, $V(\cdot)$ is a known variance function, and
	$\phi$ is a scale parameter. QLMs are also special cases of model \eqref{b11} when we select $g=\mu^{-1}$.    

Generalized models, such as generalized mean regression models \eqref{b11}, have outstanding performance in real applications.
In many real applications, determining the form of conditional variance $V(\cdot)$ is a concern. Researchers often predertermine the specific forms on the basis of prior experience. For example, \cite{NP1987} used QLMs to fit textile data. \cite{NP1987} also considered the exponential form $V(t)=t^{\alpha}$ and showed that an optimal value of $\alpha$ exists. The optimal QLM built upon the exponential form was correctly specified. The use of QLMs  ($V(t)=t+t^2/\alpha$) in insurance applications has elicited much attention. Researchers commonly use the aforementioned polynomial forms to model the number of events, such as losses to the insured or claims to the insurer. However, different forms of condition variance may lead to different variable selection and estimation results, which implies that prior-knowledge-based QLMs are not roubst in real practice. We eliminate the process of modeling conditional variance to improve the flexibility of modeling and increase
the robustness of model applications.

Suppose we have a random sample $\{Y_i,\bx_i, i=1,\ldots,n\}$ generated from model \eqref{b11}
denoted by $\bY=(Y_1,\ldots,Y_n)^\top$ and $\bX=(\bx_{1},\ldots,\bx_n)^\top$.  
Recognizing the susceptibility of existing model selection methods to the substantial intercorrelation among the components of $\bx$, we aim to decorrelate the predictor vector to mitigate the effects of correlation. 
Motivated by the  asymptotically mutual orthogonality of the features in $\bA\bX$ (\citealp{FL08,WL16}), where $\bA=(p_n^{-1}\bX\bX^\top+\lambda_n\bI_n)^{-1/2}$  and $\bI_{n}$ is an $n-$dimensional identity matrix, 
we map $(\bY,\bX)$ to $(\bA\bY,\bA\bX)$ for a linear regression model. We then conduct forward regression on $(\bA\bY,\bA\bX)$, a method referred to as decorrelated forward (DF) regression. Different from classical forward regression methods \citep{W09, CHZ16}, DF employs a decorrelated linear transformation matrix $\mathbf{A}$ to mitigate substantial intercorrelation. However, although this transformation diminishes intercorrelation, the resulting observations $(\mathbf{A}\mathbf{Y},\mathbf{A}\mathbf{X})$ are not independent, and random errors exhibit heteroscedasticity, posing a considerable challenge to demonstrating the sure screening consistency of DF. Additionally, constructing decorrelated observations for model \eqref{b11} is a formidable task. We address this challenge by transforming model \eqref{b11} into a linear regression model (\S \ref{sec2}) and employing the DF method.  Theoretically, we establish the sure screening consistency of DF under model  \eqref{b11} and provide an upper bound on  the selected submodel's size. However, the forward-searching process increases the computational burden. To devise an efficient forward regression algorithm, we propose a thresholding decorrelated forward (T-DF) algorithm that provides a stopping rule for the forward-searching process.
 
The major contributions of this work are briefly summarized here. (i) We propose a novel DF method that is robust against correlation.
(ii) We devise a novel method to transform model \eqref{b11} into a linear model (\S \ref{sec2}) and extend the forward regression idea to  model \eqref{b11}.
(iii) We establish the sure screening consistency of DF under the generalization of GLMs and QLMs, and develop a stopping rule for the forward searching process. (iv) Our DF bridges classical and popular screening methods. Figure \ref{fig1} illustrates the relationship among SIS \citep{FL08}, high-dimensional ordinary least squares projection (HOLP; \citealp{WL16}), FR \citep{W09,CHZ16}, and DF.
\tikzset{
	arrow1/.style = {
		draw =black,dashed, -{Stealth[length = 3mm, width = 2mm]},
	}	
}

\tikzset{
	arrow2/.style = {
		draw =black, semithick,densely dashed,-{Stealth[length = 3mm, width = 2mm]},
	}	
}
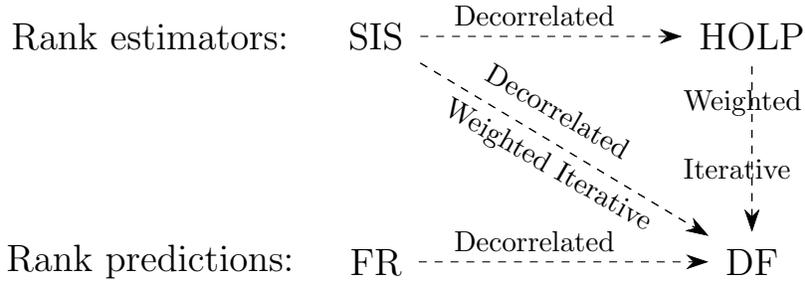
\begin{figure}[h]
	\center
	\caption{Relationship among SIS, HOLP, FR, and DF}
	\label{fig1}
	\begin{tikzpicture}[]
		
		%\draw[] (-7 ,-7);%grid ( 7, 7); 	

		\node[on grid]at(-3,2){\large Rank estimators:};
		\node[on grid]at(-3,-1){\large Rank predictions: };%$y_i$};

		\node[ rectangle,inner sep=0.22cm,label distance = 0.05cm]at(0,2) (a1){\large SIS};
		\node[right of=a1,rectangle,inner sep=0.22cm,node distance=5cm,on grid](a2) {\large HOLP};
		\node[below of=a1,rectangle,inner sep=0.22cm,node distance=3cm,on grid](b1) {\large FR};
		\node[right of=b1,rectangle,inner sep=0.22cm,node distance=5cm,on grid](b2) { \large DF};
		
		\draw[arrow1](a1)node[above,xshift = 60pt]{\small Decorrelated} -- (a2);
		\draw[arrow1](b1)node[above,xshift = 60pt]{\small Decorrelated} -- (b2);
		\draw[arrow1](a2)node[right,xshift = -30pt,yshift = -25pt]{ \small Weighted  } -- (b2);	
		\draw[arrow1](a2)node[right,xshift = -30pt,yshift = -50pt]{\small Iterative } -- (b2);
		\draw[arrow1](a1)node[above,xshift = 65pt,yshift = -35pt, rotate=-30]{\small Decorrelated} -- (b2);
		\draw[arrow1](a1)node[xshift = 65pt,yshift = -49pt, rotate=-30]{\small{Weighted  Iterative}} -- (b2);
	\end{tikzpicture}
\end{figure}

\subsection{Related literature}
\begin{figure}[hb]
	\centering
	\includegraphics[height=9cm,width=16cm]{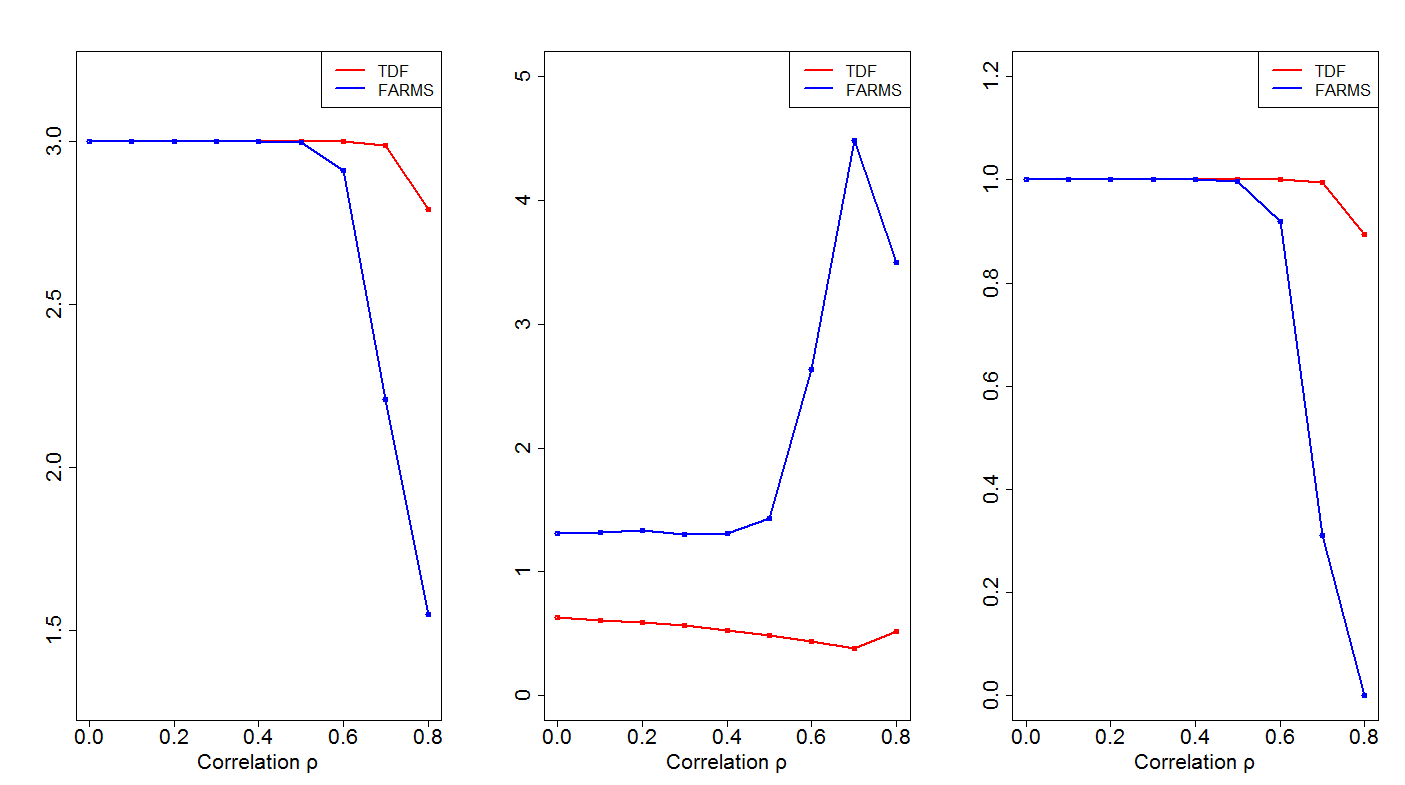}
	\vspace{-0.6cm}
	\caption{\label{toy_example}Model selection results against the correlations. The left panel shows 
		the average  number of true variables  contained in the selected model. The middle panel presents the average  number of unimportant variables mistakenly selected. The right panel shows the rate of model screening consistency.
	}
\end{figure}

To the best of our knowledge, the literature on variable selection  for model \eqref{b11} is limited because of the challenges of high dimensions and few model assumptions.
Many model selection methods are available for GLMs and QLMs.  Investigating the model selection method for model \eqref{b11} is challenging.
Under a strong irrepresentable condition \citep{BV11}, \cite{TR06} and \cite{ZY06} established the model selection consistency of LASSO, and \cite{FL11} and \cite{Sel19} extended this consistency to generalized penalty losses such as SCAD and MCP. Extensive research has been conducted on model selection consistency for penalized regression methods (\citealp{FP04, Z10, Wel23} %ZH05,
and references therein). To overcome the strong irrepresentable condition, \cite{FKW20} adopted the factor-adjusted method and proposed the factor-adjusted regularized model selection (FARMS) method. To establish model selection consistency,  \cite{FKW20} assumed that 
$\{\bx_i\}_{i=1}^n$ follows the approximate factor model
$\bx_i=\bB\mf_i+\bu_i,$
where $\{\mf_i\}_{i=1}^n \subset \mathbb{R}^{K}$ are latent factors ($K$ is independent of sample size), $\bB \in \mathbb{R}^{p_n\times K}$ is a loading matrix,  and $\{\bu_i\}_{i=1}^n\subset \mathbb{R}^{p_n}$ are idiosyncratic components. However, in practice, $K$ may diverge from the sample size. The following example shows the limitation of FARMS. Consider a special linear model  $y_i=\bx_i^\top\bbeta+\epsilon_i$, $i=1,\ldots,n$, where sample size $n=200$, dimensionality $p_n =1,000$, $\epsilon_i\sim N(0,1)$, and $\bbeta=(1,-1,0.8,0_{p_n-3})^\top$. We consider the special structure 
$\bx_i=\bB\mf_i+\bu_i$, where $\mf_i\sim N(\b0,\bI_{p_n})$, $\bu_i\sim U(0,0.5)$, and $\bB^2=(\rho^{|i-j|})_{i,j=1}^{p_n}$. In this structure, $K$ is equal to dimensionality $p_n$. 
Let $\rho $ increase from $0$ to $0.8$ by a step size of $0.1$. For each given $\rho$, we simulate $600$ replications and calculate the average  number of true variables  contained in the selected model,
the average  number of unimportant variables mistakenly selected, and the rate of model screening consistency. As shown in Figure \ref{toy_example}, as $\rho$ increases, the correlation heavily influences the model selection results of FARMS. By comparison, our DF method is robust to the substantial intercorrelation among predictors. This feature implies that our DF method is free of
the strong irrepresentable condition and robust to the complex factor structure of predictors. 

Although penalized regression methods have a good theoretical guarantee on model selection, the model selection results heavily depend on the choice of tuning parameter, leading to a high computation burden during large-scale optimization \citep{FL08}.
Additionally, \cite{FL08} introduced the SIS method for linear regression models, and it has also been extended to GLMs \citep{FSW09, FS10}. However, SIS relies on marginal correlations between the response and features, which may not hold in practice. Various forward variable screening methods have been developed to mitigate the reliance on marginal correlation assumptions. For instance,  \cite{W09} introduced the FR method, which screens variables for linear models by ranking prediction errors. Similarly, \cite{WL16} proposed the HOLP method, which screens variables by ranking estimators.

To establish screening consistency, researchers have added some restrictions on the correlation between features.	For example, \cite{FL08} assumed that the marginal correlation between important variables $x_{j}$ and $\bx^\top\bbeta^*$ is bounded away from zero. \cite{W09} assumed that the eigenvalues of the covariance matrix are  bounded away from zero and $\infty$. \cite{WL16} assumed that	$\max_{j,k\neq i}|\Omega_{ij}-\Omega_{i,k}|\to 0$ and $\max_{j,k}|\Omega_{jj}-\Omega_{kk}|\to 0$, where
$\bOmega=(E\bx\bx^\top)^{-1}$ and $\Omega_{i,k}$ is the entry of $\bOmega$ in the $i$th row and $k$th column (this is a correction to \cite{WL16}).	However, these assumptions may be violated because of the strong correlations among features in $\bx$ in high-dimensional settings.  Compared with these screening methods, our DF
relaxes the restriction on the correlations between features in $\bx$ and exhibits outstanding performance when  confronted with
substantially correlated variables in high-dimensional settings.

\subsection{Organization and notation} 
The remainder of this paper is organized as follows. In Section \ref{sec2}, we introduce our DF screening procedure and establish the screening consistency of DF and the upper bound of the size of the selected model. In Section \ref{sec5},  we
study the finite-sample performance of our approach via extensive
simulation and compare our proposed method with existing ones.
In Section \ref{sec:real}, we use two real data sets, namely, the \textit{Arabidopsis thaliana} gene dataset and the breast cancer dataset, to further demonstrate the advantages of our method.  All technical proofs are provided in the Supplementary material of \cite{Jel24}.

Throughout this paper, we adopt the following notations. For any constants $a$, $[a]$ denotes the integer part of $a$. For any vectors $\ba=(a_{1},\ldots,a_{m})^\top\in \mathbb{R}^m$ and $\bb=(b_{1},\ldots,b_{m})^\top\in \mathbb{R}^m$, we define $|\ba|=(|a_1|,\ldots,|a_m|)^\top$, $\Vert\ba\Vert_{1}=\sum_{i=1}^{m}|a_{i}|$, $\Vert \ba\Vert_{2}=\sqrt{\sum_{i=1}^{m}a_{i}^2}$, $|\ba|_{\infty}=\max_{1\leq i\leq m}|a_i|$, and $\ba\circ\bb=(a_1b_1,\ldots,a_mb_m)^\top$. For any subset $\mathcal{M}$ of the row index set  $\{1,\ldots,m\}$ of $\ba$, we let $|\mathcal{M}|$ represent the cardinality of $\mathcal{M}$, and denote by $\mathbf{a}_{\bldsi{\mathcal{M}}}$ the subvector of $\ba$ formed by row indexes in $\mathcal{M}$.  We denote by $\mathbb{I}(\cdot)$ the indicator function.

Let $\bE$ be a full-rank $l\times m$ matrix with $l> m$. The projection matrix formed by $\bE$ is denoted as $\bP(\bE)=\bE(\bE^{\top}\bE)^{-1}\bE^{\top}$. For $\mathcal{M}_1,\mathcal{M}_2\subseteq \{1,\ldots,m\}$, we let $\bE_{\bldsi{\mathcal{M}}_{1}}$ represent the submatrix of $\bE$ formed by column indexes in $\mathcal{M}_{1}$, and $\bE_{\bldsi{\mathcal{M}}_{1},\bldsi{\mathcal{M}}_{2}}$ represent the submatrix whose row and column index sets are in $\mathcal{M}_{1}$ and $\mathcal{M}_{2}$, respectively. We denote by $\lambda_{\max}(\cdot)$ and $\lambda_{\min}(\cdot)$ the maximum and minimum eigenvalues, respectively. We let $\|\bE\|_{2}=\sqrt{\lambda_{\max}(\bE^\top\bE)}$.

\section{Methodology}\label{sec2}

In this section, we propose a novel approach to convert model \eqref{b11} into a linear model.
Let  $\mathcal{Y}$ be the range of response $Y$. Assuming that inverse function $g(\cdot)$ exists within $\mathcal{Y}$, 
we obtain the expression
$$g(Y)=\bx^\top\bbeta^*+g(E(Y|\bx)+\epsilon)-g(E(Y|\bx))$$
by noting $g(E(Y|\bx))=\bx^\top\bbeta^*$ and $Y=E(Y|\bx)+\epsilon$.
This formulation yields a linear combination of covariates. Inspired by this formulation, we naturally consider the following linear regression models:
\begin{equation}\label{S2.1}
	g(Y)=\bx^\top\bbeta^*+\epsilon^*,
\end{equation}
where $\epsilon^*=g(E(Y|\bx)+\epsilon)-g(E(Y|\bx))$ denotes a random error term.

However, in a broad context, $g(\cdot)$  may not always exist within $\mathcal{Y}$.  For instance, in the Poisson model where $g(\cdot)=\log(\cdot)$, $\log(Y)$ is undefined when $Y=0$. To address this issue, we introduce the projection of  response variable $Y$, which is denoted as $\Pi_{\mathcal{Y}^*}(Y)$ and defined as
$$\Pi_{\mathcal{Y}^*}(Y)=\arg\min_{t\in\mathcal{Y}^* }|Y-t|,$$
where $\mathcal{Y}^*$ represents a carefully chosen subset of $\mathcal{Y}$ to ensure the proper definition of $g(\cdot)$.  The assumption that $\mathcal{Y^*}\subset\bar{\mathcal{R}}$, where $\bar{\mathcal{R}}$ denotes the closure of domain $\mathcal{R}$ of $g(\cdot)$, guarantees the existence of
$\Pi_{\mathcal{Y}^*}(Y)$. This assumption  is mild/versatile and applicable across various regression models, including linear, logistic, and Poisson regression.
 
To show the selection of  $\mathcal{Y}^*$, we consider a Poisson model with $g(\cdot)=\log(\cdot)$, $\mathcal{Y}=(0,\infty)$, and $\mathcal{Y}^*=[n^{-1/2},\infty)$. This choice results in  $\Pi_{\mathcal{Y}^*}(Y)=n^{-1/2}$ if $Y=0$ and $\Pi_{\mathcal{Y}^*}(Y)=Y$ otherwise.

Given the observations $\{Y_i,\bx_i\}^n_{i=1}$ from Model \eqref{b11}, 
let $\bY^*=\{Y^*_{1},\cdots,Y^*_{n}\}^\top$, where $Y^*_{i}=g(\Pi_{\mathcal{Y}^*}(Y_{i}))$
and $\bepsilon^*=(\epsilon_1^*,\ldots,\epsilon_n^*)^\top$, with $\epsilon_i^*=
g(Y_i^*)-g(E(Y_i|\bx_i))$. Then, model \eqref{S2.1} can be  presented in a sample  form as 
\begin{align}\label{GLM1}
	\bY^*=\bX\bbeta^*+\bepsilon^*,
\end{align}
where $\bX=(\bx_1,\bx_2,\cdots,\bx_n)^\top$ and $\bepsilon^*$ is a random error vector. 
Here, we allow $p_n$ to grow exponentially with $n$, such that $\log (p_n)=O(n^{\alpha})$, with $0<\alpha<1$,
and define the important index set $\mathcal{S}=\{j:\, \beta_j^*\neq 0, j=1,\cdots,p_n\}$. 

{ Since the new errors $\epsilon^*$ in \eqref{S2.1} is a nonlinear function of $\bx$,  the assumption that errors and covariates are not correlated in model $(\ref{b11})$ may not hold in model $(\ref{GLM1}) $, i.e, $E(\epsilon^*|\bx)=0$ does not always hold. However, the following result ensures consistency of the least squares estimation of model \eqref{GLM1}.
	
\begin{Proposition}{\rm \label{Proposition}
Assume Conditions $(A_4)$ and $(A_5)$ in Section \ref{ADS} hold. Then there exists some positive constant $c$ 
(dependent of function $g$) 
such that 
$$|E( x_{ij}\epsilon_i^*)| \leq cn^{-1/2}\log^2(n)\ \ \text{and} \ \  E(|x_{ij}\epsilon_i^*|^2)\leq c\log^2(n),$$
uniformly hold for $j=1,\ldots,p_n$.
Furthermore, if $g(\cdot)$ is the identity function, i.e. $\epsilon_i^*=\epsilon_i$, then 
$E(x_{ij}\epsilon_i^*)=0$ and
$E(|x_{ij}\epsilon_i^*|^2)\leq c$,
uniformly hold for $j=1,\ldots,p_n$.
}
\end{Proposition}

}

\subsection{DF regression}
{
By the singular value decomposition, there exists an $n\times n$ orthogonal matrix $\bU$,  a $p_n\times p_n$ orthogonal matrix $\bV$ and an $n\times p_n$ matrix $\bD$ such that 
\[\bX=\bU\bD\bV \ \ \text{and} \ \ (p_n^{-1}\bX\bX^\top+\lambda_{n}\bI_n)^{-1/2}\bX=p_n^{1/2} \bU(\bD\bD^\top+p_n\lambda_{n}\bI_n)^{-1/2}\bD\bV,\]
where $\lambda_{n}$ is a positive sequence that converges to zero as $n \to \infty$.
Notice that $p_n\bD^\top(\bD\bD^\top+p_n\lambda_{n}\bI_n)^{-1}\bD=\text{diag}(a_{1},a_2,\ldots,a_{p_n})$,
where $a_i=p_nd_{i}/(\lambda_{n}p_n+d_i)$ with $d_i$ being the $i$th eigenvalue value of $\bX^\top\bX$. It is obvious that 
new features in $n^{-1/2}(p_n^{-1}\bX\bX^\top+\lambda_{n}\bI_n)^{-1/2}\bX$ are approximately mutually orthogonal when $\lambda_{n}=o(\min\limits_{1\leq i\leq \min\{n,p_n\}}d_i^2/p_n^2)$ for $p_n\leq n$. Following Lemma 2 in the supplementary material of \cite{Jel24}, this property also holds for high dimensional 
settings ($p_n> n$).%(\citealp{FL08}, \citealp{WL16})
\ This motivates us to transform $\bX$ into $\mathbb{X}$:
$\mathbb{X}=(p_n^{-1}\bX\bX^\top+\lambda_{n}\bI_n)^{-1/2}\bX$, in order to decorrelate features and address the strong correlations between features.  }   
 This transformation converts model \eqref{GLM1} into
\begin{equation}\label{S2.1a}
	\mathbb{Y}^*=\mathbb{X}\bbeta^*+ \tilde\bepsilon^*,
\end{equation}
where $ \mathbb{Y}^*=(\bX\bX^\top/p_n)^{\frac{+}{2}}\bY^*$ and $\tilde\bepsilon^*=(\bX\bX^\top/p_n)^{\frac{+}{2}}\bepsilon^*$.

Next, we employ the decorrelated sum of square residuals (DRSS) to measure the importance of $\mathbb{X}_{j}$.
Given any  index $\mathcal{D}\cup\{j\}$ of $\mathbb{X}$ in Model \eqref{S2.1a}, DRSS  is defined as
\begin{eqnarray}\label{MLR}
	DR_{j}(\mathcal{D})=\|\mathbb{Y}^*-\bP(\mathbb{X}_{\bldsi{\mathcal{D}\cup\{j\}}})\mathbb{Y}^*\|_{2}^2.
\end{eqnarray}
To screen out important index set $\mathcal{S}$, we propose the DF algorithm. The details of the algorithm are listed in Algorithm 1.
Similar to the classical results of the sum of square residuals, a small value of $DR_{j}$ suggests that $\mathbb{X}_{j}$ is an important predictor.
In Step 2.2 of Algorithm 1, we add the predictor corresponding to the smallest DRSS to the candidate model.

\begin{table}
	\small%\footnotesize
	\centering
	\begin{tabular}{l}
		\toprule
		\textbf{Algorithm 1: DF algorithm} \\\hline
		\textbf{Initialization:} Input $(\bX,\bY^*)$, $\hat{\mathcal{S}}_{0}=\emptyset$,
		$n$, $\lambda_n$ and $p_n$.\\
		\textbf{(i): Decorrelated Transformation}\\
		\quad 1: Calculate $\bPsi=(\bX\bX^\top/p_n +\lambda_{n}\bI_n)^{-1/2}$, $\mathbb{X}=\bPsi\bX$ and $\mathbb{Y}^*=\bPsi
		\bY^*$;\\
		\textbf{(ii): Forward selection}\\
		\quad 2:
		For $k=1,\ldots,n$, do\\
		\qquad 2.1: Compute \{$DR_{j}(\hat{\mathcal{S}}_{k-1}):\,j\in\mathcal{S}_{k-1}^c\}$ and $j^*=\arg\min_{\bldsi{j\in\mathcal{S}_{k-1}^c}}DR_{j}(\hat{\mathcal{S}}_{k-1})$
		;\\
		\qquad 2.2: Update $\hat{\mathcal{S}}_{k}=\hat{\mathcal{S}}_{k-1}\cup\{j^*\}$.\\
		\bottomrule
	\end{tabular}
\end{table}
To illustrate our motivation, we begin with the initial search. Starting from $\hat{\mathcal{S}}_{0}=\emptyset$, we aim to obtain $\hat{\mathcal{S}}_{1}$. 
In Step 2.1 of the DF algorithm, we rank $DR_{j}(\hat{\mathcal{S}}_{0})$ and choose the index corresponding to the smallest DRSS. On the basis of the definition of DRSS in (\ref{MLR}), we have 
\begin{align}\label{Drc}
	DR_{j}(\hat{\mathcal{S}}_{0})&=\|\mathbb{Y}^*\|_{2}^2-\|\bP(\mathbb{X}_{j})\mathbb{Y}^*\|_{2}^2\nonumber\\
	%&=\|\mathbb{Y}^*\|_{2}^2- \|\bPsi\bX_{j}(\bX_{j}^\top\bPsi^2\bX_{j})^{-1}\bX_{j}^\top\bPsi^2\bY^*\|_{2}^2\nonumber\\
	&=\|\mathbb{Y}^*\|_{2}^2- |(\bX_{j}^\top\bPsi^2\bX_{j})^{-1/2}\bX_{j}^\top\bPsi^2\bY^*|^2.
\end{align}
Thus, $\min_{j}DR_{j}(\hat{\mathcal{S}}_{0})$ is equivalent to $\max_{j}|(\bX_{j}^\top\bPsi^2\bX_{j})^{-1/2}\bX_{j}^\top\bPsi^2\bY^*|$.
The ridge-HOLP \citep{WL16} and SIS \citep{FL08} estimators for Model (\ref{GLM1}) are
$$\hat\bbeta^{\bldsr {RH}}=\bX^\top\bPsi^2\bY^*, \ \ \hat\bbeta^{\bldsr {SIS}}=\bX^\top\bY^*.$$
We define the weighted ridge-HOLP estimator as
$$
\hat\bbeta^{\bldsr {WRH}}=\bU\bX^\top\bPsi^2\bY^*,
$$
where $\bU=\text{diag}\{(\bX_{1}^\top\bPsi^2\bX_{1})^{-1/2},\ldots, (\bX_{p}^\top\bPsi^2\bX_{p})^{-1/2}\}$.
These formulas show that the weighted ridge-HOLP estimator $\hat\bbeta^{\bldsr{WRH}}$ is also the weighted decorrelated SIS estimator obtained by applying $\hat\bbeta^{\bldsr {SIS}}$ to the transformed Model (\ref{S2.1a}).  While HOLP and SIS directly rank estimators to obtain the important index set, our DF approach adopts iterative procedures and adds one predictor at each step. DF can be viewed as an iterative weighted ridge-HOLP approach (or decorrelated SIS), that aims to mitigate correlation effects. The weighted operator, as shown in \eqref{Drc}, accounts for  the marginal prediction contribution. Unlike SIS and HOLP that only consider estimator magnitudes, our DF approach utilizes the full contribution of covariates to predictions.

\subsection{Asymptotics of DF screening}\label{ADS}

In this section, we establish the asymptotic properties of our estimator. We introduce some conditions into the likelihood  and penalty functions and present the theoretical results.

Let $\sigma_n^2=E|Y_{1}^*|^2$, $\bSigma=E(\mathbf{xx^{\top}})$, $\mathbf{W=X}{\bSigma}^{-1/2}$, and $\mathbf{w}={\bSigma}^{-1/2}\mathbf{x}$. The following conditions are required for screening.

{\spacingset{1.5}
\begin{itemize}
	\item[($A_1$)] The random vector $\bw$  has a spherically symmetric distribution, and
	constants $c_{1},C_{1}>0$ exist, such that
	\[P(\lambda_{\max}(p_{n}^{-1}\mathbf{WW^{\top}})>c_{1},\ \ \text {or} \ \ \lambda_{\min}(p_{n}^{-1}\mathbf{WW^{\top}})<1/c_{1})\leq{\exp(-C_{1}n)}.\]
	\item[($A_2$)] For some $\kappa>{0}$, $\tau\geq{0}$, and $c_{2},c_{3},c_{4}>0$,
	\[\min_{j\in{\mathcal{S}}}|\beta_{j}^*|\geq\frac{c_{2}}{n^{\kappa}},  \ \ \text{and} \ \
	c_{3}n^{-\tau/2}\leq \lambda_{\min}(\bSigma) \leq \lambda_{\max}(\bSigma)\leq c_{4}n^{\tau/2}.\]
	\item[($A_3$)] $\{s_{n}+n^{4.5\tau+2\kappa}\sigma_n^2\}^{1/2}n^{5\tau+2\kappa}\log^2(n)\log(p_n)=o(n)$,
	$\{s_{n}+n^{4.5\tau+2\kappa}\sigma_n^2\}n^{3\tau/4}$
	$\log(p_n)=o(n^{1/2})$ and $n^{3\tau/2}\lambda_{n}=o(1)$.
	\item[$(A_{4})$] Let $\xi_{i}$ be a constant satisfying  $g(\Pi_{\mathcal{Y}^*}(Y_i))-\bx_{i}^\top\bbeta^*=\xi_{i}\{\Pi_{\mathcal{Y}^*}(Y_i)-
	g^{-1}(\bx_{i}^\top\bbeta^*)\}$.
	The range $\mathcal{Y}^*$ and  constant $c_{5}>0$ exist, such that with probability approaching one,%\sigma_n^2\leq  c_{5}\log(n),
	\[ \ \ |\Pi_{\mathcal{Y}^*}(Y_i)-Y_{i}|\leq c_{5}n^{-1/2}\ \ \text{and} \ \ |\xi_{i}|< c_{5}\log(n)\]
	holds uniformly.
	\item[$(A_{5})$] Constants $c_6,c_{7}>0$ exist, such that 
	%	At least one of the following two conditions holds:
	%	\begin{itemize}
	%		\item [(i)]
	%		The function $b''(z)$ is positive and bounded in its domain: $b''(z)\leq c_{6}$ for some constant $c_{6}>0$;
	%		\item [(ii)]
	
	(a) $\max_{1\leq i\leq n} E\big\{\exp(c_{6}|\varepsilon_{i}|)-1-c_{6}|\varepsilon_{i}|\mid\bx_i\}\leq \frac{c_{6}^2 c_{7}}{2}$;
	
	(b)
	$\|\ba^\top\bx_{1}\|_{\psi_{2}}\leq c_{6}\|\ba\|_{2}$ for any $\ba\in R^{p_n}$,
	where $\|\ba^\top\bx_{1}\|_{\psi_{2}}=\inf_{m>0}\{E\exp(m^{-2}|\ba^\top\bx_{1}|^2)\leq 2\}$.
	%	\end{itemize}
\end{itemize}
}
{
Conditions $(A_1)$ holds when $\bx$ is $p_n-$dimensional Gaussian distributed (see \cite{FL08}). The deviation inequality in Condition $(A_1)$ is also true for any sub-Gaussian distributions (see \citealp{WL16}).
Condition $(A_1)$ also holds for a wide class of spherically symmetric distributions (see \cite{FL08} for details). Furthermore, Condition $(A_1)$ was assumed by \cite{FL08} and \cite{WL16}.
In Condition $(A_2)$, $\kappa$ controls the strength of signals, and $\tau$ controls the singularity of covariance matrix. Condition $(A_3)$ restricts the diverging rate of the dimensionality $p_n$.
\cite{W09} assumed  $\tau=0$ and $\log(p_{n})s_{n}^6n^{12\kappa}=o(n)$, which is stronger than Condition $(A_3)$. As for $\lambda_{n}$, we can choose $\lambda_{n}=c(\log(p_n)/n)^{1/4}$ for some constant $c$ in practice, which guarantee $\lambda_{n}n^{1.5\tau}=o(1)$.  
}

{\spacingset{1.5}
For Condition $(A_4)$, we demonstrate the selection of $\mathcal{Y}^*$ for linear, logistic, and Poisson regression models.
\begin{itemize}
	\item [(i)]For linear regression models, we set $\mathcal{Y}^*=R$. Then, $\Pi_{\mathcal{Y}^*}(Y_i)=Y_{i}$.
	\item [(ii)] For logistic regression models, we set $\mathcal{Y}^*=[n^{-1/2},1-n^{-1/2}]$. Then,
	$\Pi_{\mathcal{Y}^*}(Y_i)=n^{-1/2}\mathbb{I}(Y_{i}=0)+(1-n^{-1/2})\mathbb{I}(Y_{i}=1)$.
	\item[(iii)] For Poisson regression models, we set $\mathcal{Y}^*=[n^{-1/2},\infty)$. Then,
	$\Pi_{\mathcal{Y}^*}(Y_i)=n^{-1/2}\mathbb{I}(Y_{i}=0)+Y_{i}\mathbb{I}(Y_{i}\neq 0)$.
	
	\item [(iv)] For \text{mean regression models with $g^{-1}(t)=t^{\alpha}$ and $\alpha=1/3,1/5$},  we set $\mathcal{Y}^*=R$. Then, $\Pi_{\mathcal{Y}^*}(Y_i)=Y_{i}$.
\end{itemize}
}

The discussion on Condition $(A_{4})$ in the Supplementary material of \cite{Jel24} reveals that if constants $\delta_1$ and $\delta_2$ are greater than zero, such that $0 < \delta_{1} \leq g^{-1}(\bx_{i}^\top\bbeta^*) \leq \delta_{2} < 1$ for logistic and mean regression models and $g^{-1}(\bx_{i}^\top\bbeta^*) \geq \delta_{1} > 0$ for Poisson regression models, then Condition $(A_{4})$ holds for linear, logistic, and Poisson regression models as well as for nonlinear regression models with $g^{-1}(t)=t^{\alpha}$, where $\alpha=1/3$ or $1/5$. Furthermore, the existence of $\delta_1$ and $\delta_2$ is implied by $\max_{i}|\bx_{i}^\top\bbeta^*|=O(1)$, as assumed by \cite{Vel14}, \cite{NL17}, and \cite{Sel21}.

Condition $(A_5)$(a) holds for the Gaussian linear regression, logistic regression, and poisson regression. Condition $(A_5)$(a) is also true for the sub-Gaussian errors. Condition $(A_{5})$(a) was also assumed by \cite{FL11}. Condition $(A_5)$(b) needs $\bx_{1}$ to be a sub-Gaussian vector, which was assumed by \cite{Vel14}, \cite{NL17}, and \cite{Sel21}.

To screen the variables, we define $\hat{\mathcal{S}}_{k_{0}}$ as the smallest length set that contains $\mathcal{S}$, such that  $\mathcal{S}\not\subset \hat{\mathcal{S}}_{k_{0}-1}$. According to the definition of $\hat{\mathcal{S}}_k$ in Algorithm 1, $\hat{\mathcal{S}}_{k_{0}}$ is not an empty set.
The following theorem characterizes the existence of $\hat{\mathcal{S}}_{k_{0}}$ and provides an upper bound for its size.
\begin{Theorem}\label{th1}
	{\rm
		Assume that Conditions ($A_1$)--($A_5$) hold. With probability approaching one, there exists a positive constant $c_{0}$ such that $$k_{0}\leq s_{n}+ c_0\sigma_n^2n^{4.5\tau+2\kappa},$$
		where $\sigma_n^2=E|Y_{i}^*|^2$, $s_{n}=|\mathcal{S}|$, and $\tau$ and $\kappa$ are defined in Condition $(A_2)$ .
}\end{Theorem}

\begin{Remark}
	{\rm
		Under Condition ($A_{2}$), Theorem \ref{th1} suggests that the upper bound of the selected model size depends on the minimum signal strength $\min_{j\in \mathcal{S}}|\beta_{j}^*|\geq c_{2}n^{-\kappa}$ and the upper bound of the conditional number $\lambda_{\max}(\bSigma)/\lambda_{\min}(\bSigma)\leq c_4 n^\tau/c_3$ of $\bSigma$, which intuitively makes sense.
	}
\end{Remark}

\begin{Remark}
	{\rm For linear regression,
		\cite{W09} demonstrated  that the upper bound of the selected model size by the FR algorithm  is $O(s_{n}^2n^{4\kappa})$ with $\tau=0$ and $\sigma_n^2=O(1)$.
		According to Theorem \ref{th1} and $\tau=0$, we find that $|\hat{\mathcal{S}}_{k_0}|\leq s_{n}+c_{0}n^{2\kappa}$. Thus, our result surpasses that of \cite{W09}.
	{
    We extended the FR to generalized mean regression models via nonlinear transformation in model \eqref{GLM1} and established that we can detect all important variables up to the size of the selected model equal to $O(s_{n}+ c_0\sigma_n^2n^{4.5\tau+2\kappa})$. Different from linear regression, our results introduce the new term $\sigma_n^2$. For linear models, $\sigma_n^2=O(1)$ when $E|\bx_{i}^\top\bbeta^*|^2=O(1)$, which is commonly used in the high dimensional regression \citep{W09,NL17,Sel19}. For model \eqref{GLM1}, $\sigma_n^2$ is allowed to be $O(\log^2(n))$, by Proposition \ref{Proposition}. 
	}
	}
\end{Remark}

To determine when to terminate the forward search in Step 2 of Algorithm 1, we assess the importance of $\hat{\mathcal{S}}_{k}\setminus\hat{\mathcal{S}}_{k-1}$ by comparing the differences in DRSS. A considerable difference in DRSS indicates that the covariate corresponding to the index $\hat{\mathcal{S}}_{k}\setminus\hat{\mathcal{S}}_{k-1}$ contributes considerably to prediction. Using this observation, we introduce the thresholding decorrelated forward (T-DF) algorithm, outlined as Algorithm 2.

\begin{table}
	\small%\footnotesize
	\centering
	\begin{tabular}{l}
		\toprule
		\textbf{Algorithm 2 T-DF algorithm} \\\hline
		\textbf{Initialization:} Input $(\bX,\bY^*)$, $\hat{\mathcal{S}}_{0}=\emptyset$, $c$,
		$n$, and $p_n$.\\
		\textbf{(i): Decorrelated Transformation}\\
		\quad 1: Calculate $\bPsi=(\bX\bX^\top/p_n +\lambda_{n}\bI_n)^{-1/2}$, $\mathbb{X}=\bPsi\bX$ and $\mathbb{Y}^*=\bPsi
		\bY^*$;\\
		\textbf{(ii): Forward selection}\\
		\quad 2:
		For $k=1,\ldots,n$, do\\
		\qquad 2.1: Compute \{$DR_{j}(\hat{\mathcal{S}}_{k-1}):\,j\in\mathcal{S}_{k-1}^c\}$ and $j^*=\arg\min_{j\in\mathcal{S}_{k-1}^c} DR_{j}(\hat{\mathcal{S}}_{k-1})$
		;\\
		\qquad 2.2: Update $\hat{\mathcal{S}}_{k}=\hat{\mathcal{S}}_{k-1}\cup\{j^*\}$.\\
		\qquad 2.3: Let $c_{n,k}=ck\|\bPsi\|_2^2\log(\log(n^{1/3}))\log(p_n)\{\mathbb{I}(\bY=\bY^*)+\sqrt{\log(p_n)}\mathbb{I}(\bY\neq\bY^*)\}$;\\
		\qquad 2.4: Compute $\ell_{k}=
		\|\mathbb{Y}^*-\bP(\mathbb{X}_{\hat{\mathcal{S}}_{k-1}})\mathbb{Y}^*\|_{2}^2-\|\mathbb{Y}^*-\bP(\mathbb{X}_{\hat{\mathcal{S}}_{k}})\mathbb{Y}^*\|_{2}^2
		$;\\
		\qquad 2.4: If   $\ell_{k}\leq c_{n,k}$, stop and update $\hat{\mathcal{S}}=\hat{\mathcal{S}}_{k-1}$.\\
		\quad 3: Return $\hat{\mathcal{S}}$.\\
		\bottomrule
	\end{tabular}
\end{table}

The following theorem establishes the screening consistency of the T-DF algorithm.
\begin{Theorem}\label{th2}
	{\rm
		Assume that Conditions ($A_1$)--($A_5$) hold. Then,
		$$P(\hat{\mathcal{S}}=\hat{\mathcal{S}}_{k_0})\to 1.$$
}\end{Theorem}
Theorem \ref{th2} states that a data-driven true submodel $\hat{\mathcal{S}}_{k_0}$ with probability approaching one is selected. Additionally,  
the screening consistency of the T-DF algorithm is established, and an upper bound for the size of the selected model is provided. {
In settings of strong correlation, classical regularization methods and correlation-based model selection approaches (such as SIS, FR, and HOLP) lack theoretical guarantees. However, Theorem \ref{th2} ensures that our T-DF algorithm also achieves weak model selection consistency.
}

\begin{Remark}{\rm
		In the context of Theorems \ref{th1} and \ref{th2}, we observe that the linear transformation described by Model \eqref{GLM1} remarkably facilitates model selection and provides a novel approach to perform dimension reduction of covariates for complex datasets with a limited  model structure.   Although the least squares estimator associated with Model \eqref{GLM1} lacks certain desirable statistical properties inherent to the maximum likelihood estimator, such as unbiasedness and asymptotic normality, it has a strong screening consistency.  This property enables us to distinguish the important variables from the noise variables with probability approaching one. Consequently, our proposed transformation approach provides a useful algorithmic framework for effective variable selection.
	}
\end{Remark}

\section{Numerical experiments}\label{sec5}
In this section, we present the screening results of the proposed T-DF method and compare them with the results of several existing model selection methods, including LASSO \citep{T96}, SCAD \citep{FL01,FL11}, iterative SIS (I-SIS; \citealp{FSW09}) , FR \citep{W09},  HOLP \citep{WL16}, C-FS \citep{Z20}, and FARMS \citep{FKW20}.  Notably, HOLP \citep{WL16} and FR \citep{W09}  are only applicable to {linear models (Example \ref{lin})}.  To determine parameter $c$ in Step 2.3 of the T-DF algorithm, we employ 10-fold cross validation and select $c$ to minimize the {average} prediction error. { The orthogonality parameter is set as $\lambda_{n}=4\{\log(p_n)/n\}^{1/4}$ so that new features in $n^{-1/2}(p_n^{-1}\bX\bX^\top+\lambda_{n}\bI_n)^{-1/2}\bX$ are approximately mutually  orthogonal.}
 The tuning parameters for LASSO \citep{T96} and SCAD \citep{FL01,FL11} are chosen using the Bayesian information criterion (BIC; \citealp{Sch78}).  Additionally,  we utilize the R Package  \texttt{SIS} for I-SIS \citep{FSW09} and apply the vanilla version of I-SIS. The penalized subproblems in I-SIS \citep{FSW09} adopt the SCAD penalty function, and the tuning parameters are determined via BIC \citep{Sch78}.
Moreover, we choose the optimal candidate model in HOLP \citep{WL16} and FR \citep{W09} by utilizing the extended BIC \citep{CC08}. 
We implement FARMS \citep{FKW20} by using R Package \texttt{FarmSelect}.

We evaluate the performance of the aforementioned methods by adopting the following metrics:  (1) true positive (TP), which represents the number of true variables correctly identified; (2) false positive (FP), which indicates the count of unimportant variables mistakenly selected; (3) {coverage ratio (CR)}, that denotes the proportion of true variables contained in the selected model. A value of 1 signifies $\mathcal{S}\subset\hat{\mathcal{S}}$, and 0 indicates the opposite. We report the average values and standard deviations of these metrics over 600 repetitions. All algorithms are executed on a computer equipped with an Intel(R) Xeon(R) Gold 6142 CPU and 256 GB of RAM.

In each simulation, we examine the effect of feature space dimensionality.  Initially, we set $n=200$ and vary $p_n$ from $500$ to $1,000$. Subsequently, we increase the sample size to $n=400$ to further examine the effect of sample size.

Furthermore,  we investigate the performance of penalized methods in Model (\ref{S2.1a}) obtained through  decorrelation transformation. However, the results  exhibit deterioration, and for brevity, we do not report these findings.

\begin{Example}{\rm  \label{lin}%(\textit{high dimensional linear regression})
		For the linear model, we set
		\begin{equation*}
			Y_i=\bx^\top_i\bbeta^*+ \epsilon_i,\qquad i=1,\dots,n.
		\end{equation*}
	}
\end{Example}

\begin{table}[htb]
	\centering
	\scriptsize
	\caption{TP, FP, and CR ($\%$) over 600 repetitions and standard deviations for linear regression in Scenario (I) with $n=200$}
	\label{tab1}
	\vspace{-0.2cm}
	\resizebox{1\textwidth}{!}{
		\begin{tabular}{cc|ccccccccc}
			\hline
			&&TP&FP&CR&TP&FP&CR&TP&FP&CR\\ \hline
			Approach &$p_n$&\multicolumn{3}{c}{$\rho=0$}&\multicolumn{3}{c}{$\rho=0.5$}&\multicolumn{3}{c}{$\rho=0.8$}\\ 
			\hline
			\multirow{2}{*}{\textsc{T-DF}}&500 & 3.00(0.00)&0.14(0.35)&1.00(0.00)&
			3.00(0.00)&0.11(0.31)&1.00(0.00)&
			2.88(0.48)&0.20(0.50)&0.94(0.24)\\
			&1000&3.00(0.00)&0.63(0.52)&1.00(0.00)& 3.00(0.00)&0.50(0.52)&1.00(0.00)&
			2.67(0.74)&0.67(0.84)&0.83(0.38)\\ \hline
			\multirow{2}{*}{\textsc{LASSO}}&500 & 3.00(0.00)&0.26(0.52)&1.00(0.00)&
			2.89(0.34)&1.03(1.06)&0.90(0.30)&
			1.89(0.32)&0.68(1.05)&0.00(0.00)\\
			&1000&3.00(0.00)&0.62(0.82)&1.00(0.00)&
			2.84(0.43)&1.92(1.60)&0.86(0.34)&
			1.87(0.34)&0.61(0.96)&0.00(0.00)\\ \hline
			\multirow{2}{*}{\textsc{FBIC}}&500 & 3.00(0.00)&0.01(0.10)&1.00(0.00)&
			2.97(0.26)&0.02(0.13)&0.99(0.12)&
			1.27(0.67)&0.02(0.15)&0.13(0.34)\\
			&1000&3.00(0.00)&0.02(0.15)&1.00(0.00)&
			2.98(0.20)&0.01(0.10)&0.99(0.10)&
			1.12(0.47)&0.02(0.16)&0.06(0.23)\\ \hline
			\multirow{2}{*}{\textsc{SCAD}}&500 & 3.00(0.00)&0.19(0.54)&1.00(0.00)& 3.00(0.00)&0.47(1.02)&1.00(0.00)&
			1.49(0.54)&0.65(1.08)&0.02(0.15)\\
			&1000&3.00(0.00)&0.28(0.69)&1.00(0.00)&
			3.00(0.00)&0.75(1.33)&1.00(0.00)&
			1.51(0.50)&0.59(0.91)&0.00(0.00)\\ \hline
			\multirow{2}{*}{\textsc{I-SIS}}&500 & 3.00(0.00)&0.20(0.54)&1.00(0.00)&
			3.00(0.00)&0.69(2.55)&1.00(0.00)&
			2.56(0.78)&12.07(10.46)&0.75(0.44)\\
			&1000&3.00(0.00)&0.32(1.50)&1.00(0.00)&
			3.00(0.00)&1.02(2.69)&1.00(0.00)&
			2.59(0.76)&19.66(12.57)&0.76(0.43)\\ \hline
			\multirow{2}{*}{\textsc{C-FS}}&500 &3.00(0.00)&3.13(2.15)&1.00(0.00)&
			3.00(0.00)&2.74(2.01)&1.00(0.00)&
			2.55(0.83)&2.05(1.61)&0.78(0.42)\\
			&1000&3.00(0.00)&6.23(3.55)&1.00(0.00)& 3.00(0.00)&5.83(3.61)&1.00(0.00)&
			2.41(0.91)&4.09(2.64)&0.70(0.46)\\ \hline
			\multirow{2}{*}{\textsc{HOLP}}&500 & 3.00(0.00)&0.00(0.06)&1.00(0.00)&
			2.14(0.84)&0.01(0.10)&0.43(0.50)&
			1.11(0.33)&0.02(0.15)&0.00(0.00)\\
			&1000&3.00(0.00)&0.00(0.00)&1.00(0.00)& 1.53(0.59)&0.00(0.00)&0.05(0.21)&
			1.05(0.23)&0.01(0.07)&0.00(0.00)\\ \hline
			\multirow{2}{*}{\textsc{FARMS}}&500 & 3.00(0.00)&1.27(1.17)&1.00(0.00)&
			3.00(0.00)&1.44(1.44)&1.00(0.00)&
			1.52(0.50)&2.77(1.83)&0.00(0.00)\\
			&1000&3.00(0.00)&1.36(1.31)&1.00(0.00)& 3.00(0.00)&1.62(1.78)&1.00(0.00)&
			1.51(0.50)&3.27(2.07)&0.00(0.00)\\
			\hline
		\end{tabular}
	}
\end{table}
We consider two correlation structures of $\bx_{i}$: 
(I) autoregressive correlation:  $\bx_{i}=\bSigma^{1/2}\mf_i$,
where $\bSigma=(\rho^{|i-j|})_{i,j=1}^{p_n}$ is a positive definite matrix and 
$\mf_i$s are independently identically distributed ($i.i.d$) from a multivariate normal distribution with mean $\b0$ and covariance matrix $\bI_{p_n}$;
(II) block compound symmetry correlation: 
$\bx_i=\bSigma^{1/2}\mf_i-\bu_i$,
where $\bu_i=(\b0_{3}^\top,0.6a_i\rho\mathbf{1}^\top_{p_n-3})^\top$ with
$a_i$ being the first component of $\bSigma^{1/2}\mf_i$,
and
$\bSigma$ has  diagonal elements set to 1 and off-diagonal elements set to $\rho$.
We set the true regression coefficient as $\bbeta^*=(1,-1,0.8,\textbf{0}_{p_n -3})^\top$.  Random errors $\epsilon_i$ are $i.i.d$ copies from $N(0,1)$. 
We consider three levels of $\rho$ with $\rho=0,0.5 \ {\rm or}\ 0.8$.

\begin{table}[htb]
	\centering
	\scriptsize
	\caption{TP, FP, and CR ($\%$) over 600 repetitions and standard deviations for linear regression in Scenario (II) with $n=200$}
	\label{tab2}
	\vspace{-0.2cm}
	\resizebox{0.8\textwidth}{!}{
		\begin{tabular}{cc|cccccc}
			\hline
			&&TP&FP&CR&TP&FP&CR\\ \hline
			Approach &$p_n$&\multicolumn{3}{c}{$\rho=0.5$}&\multicolumn{3}{c}{$\rho=0.8$}\\ 
			\hline
		\multirow{2}{*}{\textsc{T-DF}}&500 & 3.00(0.00)&0.13(0.34)&1.00(0.00)&
		2.91(0.39)&1.09(0.72)&0.94(0.24)
		\\
		&1000&3.00(0.00)&0.44(0.51)&1.00(0.00)& 2.83(0.51)&1.31(0.79)&0.89(0.31)
		\\ \hline
		\multirow{2}{*}{\textsc{LASSO}}&500 & 3.00(0.00)&0.60(0.84)&1.00(0.00)&
		2.17(0.48)&1.24(1.77)&0.22(0.41)
		\\
		&1000&3.00(0.00)&0.75(1.11)&1.00(0.00)&
		2.07(0.44)&1.17(2.04)&0.13(0.34)\\ \hline
		\multirow{2}{*}{\textsc{FBIC}}&500 & 3.00(0.00)&0.03(0.17)&1.00(0.00)&
		2.39(0.89)&0.14(0.35)&0.67(0.47)\\
		&1000&3.00(0.00)&0.02(0.15)&1.00(0.00)&
		2.29(0.91)&0.15(0.36)&0.61(0.49)\\ \hline
		\multirow{2}{*}{\textsc{SCAD}}&500 & 3.00(0.00)&0.23(0.57)&1.00(0.00)& 1.92(0.87)&1.02(1.83)&0.34(0.47)\\
		&1000&3.00(0.00)&0.42(0.95)&1.00(0.00)&
		1.71(0.80)&1.14(2.13)&0.22(0.41)\\ \hline
		\multirow{2}{*}{\textsc{I-SIS}}&500 &3.00(0.00)&0.31(1.53)&1.00(0.00)&
		2.85(0.44)&5.16(7.88)&0.88(0.33)\\
		&1000&3.00(0.00)&0.63(2.59)&1.00(0.00)&
		2.83(0.44)&10.59(11.54)&0.86(0.35)\\ \hline
		\multirow{2}{*}{\textsc{C-FS}}&500 &
		3.00(0.00)&2.16(1.93)&1.00(0.00)&2.80(0.55)&2.30(1.99)&0.87(0.34)\\
		&1000&3.00(0.00)&4.86(3.50)&1.00(0.00)& 2.78(0.56)&4.75(3.62)&0.85(0.36)\\ \hline
		\multirow{2}{*}{\textsc{HOLP}}&500&
		2.29(0.95)&0.01(0.11)&0.64(0.48) & 1.13(0.44)&0.01(0.10)&0.04(0.20)\\
		&1000&1.95(0.96)&0.04(0.20)&0.44(0.50)&  1.01(0.10)&0.01(0.04)&0.00(0.00)\\ \hline
		\multirow{2}{*}{\textsc{FARMS}}&500 & 3.00(0.00)&1.38(1.24)&1.00(0.00)&
		2.37(0.65)&2.45(1.70)&0.47(0.50)\\
		&1000&3.00(0.00)&1.57(1.31)&1.00(0.00)& 2.17(0.65)&3.11(2.02)&0.31(0.46)\\
			\hline
		\end{tabular}
	}
\end{table}

\begin{table}[htb]
	\centering
	\scriptsize
	\caption{TP, FP and CR ($\%$) over 600 repetitions and standard deviations for linear regression in Scenario (I) with $n=400$}
	\label{tab3}
	\vspace{-0.2cm}
	\resizebox{1\textwidth}{!}{
		%	\begin{center}
			\begin{tabular}{cc|ccccccccc}
				\hline
				&&TP&FP&CR&TP&FP&CR&TP&FP&CR\\ 
				\hline
				Approach &$p_n$&\multicolumn{3}{c}{$\rho=0$}&\multicolumn{3}{c}{$\rho=0.5$}&\multicolumn{3}{c}{$\rho=0.8$}\\ 
				\hline
				\multirow{2}{*}{\textsc{T-DF}}&500 & 3.00(0.00)&0.04(0.19)&1.00(0.00)&
				3.00(0.00)&0.15(0.35)&1.00(0.00)&
				3.00(0.00)&0.49(0.57)&1.00(0.00)\\
				&1000&3.00(0.00)&0.08(0.26)&1.00(0.00)& 3.00(0.00)&0.26(0.44)&1.00(0.00)&
				3.00(0.00)&0.65(0.56)&1.00(0.00)\\ \hline
				\multirow{2}{*}{\textsc{LASSO}}&500 & 3.00(0.00)&0.58(0.81)&1.00(0.00)&
				3.00(0.00)&0.48(0.72)&1.00(0.00)&
				2.03(0.21)&0.74(1.62)&0.04(0.19)\\
				&1000&3.00(0.00)&0.65(0.88)&1.00(0.00)&
				3.00(0.00)&0.58(0.74)&1.00(0.00)&
				1.99(0.08)&0.50(0.83)&0.00(0.00)\\ \hline
				\multirow{2}{*}{\textsc{FBIC}}&500 & 3.00(0.00)&0.01(0.09)&1.00(0.00)&
				3.00(0.00)&0.02(0.13)&1.00(0.00)&
				2.44(0.90)&0.03(0.17)&0.72(0.45)\\
				&1000&3.00(0.00)&0.01(0.11)&1.00(0.00)&
				3.00(0.00)&0.01(0.10)&1.00(0.00)&
				2.34(0.94)&0.02(0.13)&0.67(0.47)\\ \hline
				\multirow{2}{*}{\textsc{SCAD}}&500 & 3.00(0.00)&0.06(0.25)&1.00(0.00)& 3.00(0.00)&0.06(0.26)&1.00(0.00)&
				2.04(0.92)&0.71(1.19)&0.44(0.50)\\
				&1000&3.00(0.00)&0.07(0.33)&1.00(0.00)&
				3.00(0.00)&0.05(0.24)&1.00(0.00)&
				1.73(0.80)&0.48(0.97)&0.22(0.42)\\ \hline
				\multirow{2}{*}{\textsc{I-SIS}}&500 & 3.00(0.00)&0.05(0.23)&1.00(0.00)&
				3.00(0.00)&0.05(0.28)&1.00(0.00)&
				2.94(0.33)&3.60(3.34)&0.97(0.17)\\
				&1000&3.00(0.00)&0.05(0.29)&1.00(0.00)&
				3.00(0.00)&0.05(0.27)&1.00(0.00)&
				2.88(0.48)&6.01(5.37)&0.93(0.25)\\ \hline
				\multirow{2}{*}{\textsc{C-FS}}&500 &3.00(0.00)&2.75(1.94)&1.00(0.00)&
				3.00(0.00)&2.52(1.75)&1.00(0.00)&
				2.99(0.14)&1.76(1.50)&1.00(0.00)\\
				&1000&3.00(0.00)&5.58(3.06)&1.00(0.00)& 3.00(0.00)&5.11(2.95)&1.00(0.00)&
				2.99(0.12)&3.48(2.04)&1.00(0.00)\\ \hline
				\multirow{2}{*}{\textsc{HOLP}}&500 & 3.00(0.00)&0.00(0.00)&1.00(0.00)&
				2.91(0.33)&0.00(0.06)&0.92(0.27)&
				1.53(0.86)&0.10(0.31)&0.20(0.40)\\
				&1000&3.00(0.00)&0.00(0.04)&1.00(0.00)& 2.89(0.40)&0.00(0.06)&0.91(0.28)&
				1.44(0.51)&0.00(0.06)&0.01(0.09)\\ \hline
				\multirow{2}{*}{\textsc{FARMS}}&500 & 3.00(0.00)&1.33(1.23)&1.00(0.00)&
				3.00(0.00)&1.26(1.23)&1.00(0.00)&
				2.38(0.68)&2.24(1.69)&0.50(0.50)\\
				&1000&3.00(0.00)&1.47(1.28)&1.00(0.00)& 3.00(0.00)&1.31(1.18)&1.00(0.00)&
				2.03(0.66)&3.10(1.96)&0.24(0.42)\\
				\hline
			\end{tabular}
			%	\end{center}
	}
\end{table}

\begin{table}[htb]
	\centering
	\scriptsize
	\caption{TP, FP and CR ($\%$) over 600 repetitions and standard deviations for linear regression in Scenario (II) with $n=400$}
	\label{tab4}
	\vspace{-0.2cm}
	\resizebox{0.8\textwidth}{!}{
		\begin{tabular}{cc|cccccc}
			\hline
			&&TP&FP&CR&TP&FP&CR\\ 
			\hline
			Approach &$p_n$&\multicolumn{3}{c}{$\rho=0.5$}&\multicolumn{3}{c}{$\rho=0.8$}\\ 
			\hline
				\multirow{2}{*}{\textsc{T-DF}}&500 & 3.00(0.00)&0.14(0.35)&1.00(0.00)&
			3.00(0.00)&0.43(0.52)&1.00(0.00)
			\\
			&1000&3.00(0.00)&0.25(0.44)&1.00(0.00)& 3.00(0.00)&0.52(0.54)&1.00(0.00)
			\\ \hline
			\multirow{2}{*}{\textsc{LASSO}}&500 & 3.00(0.00)&0.40(0.75)&1.00(0.00)&
			2.33(0.48)&0.30(0.69)&0.33(0.47)
			\\
			&1000&3.00(0.00)&0.72(1.00)&1.00(0.00)&
			2.40(0.50)&0.59(1.04)&0.40(0.49)\\ \hline
			\multirow{2}{*}{\textsc{FBIC}}&500 & 3.00(0.00)&0.01(0.11)&1.00(0.00)&
			2.99(0.16)&0.03(0.16)&0.99(0.08)\\
			&1000&3.00(0.00)&0.02(0.13)&1.00(0.00)&
			2.98(0.19)&0.04(0.20)&0.99(0.10)\\ \hline
			\multirow{2}{*}{\textsc{SCAD}}&500 & 3.00(0.00)&0.05(0.24)&1.00(0.00)& 3.00(0.00)&0.19(0.78)&1.00(0.00)\\
			&1000&3.00(0.00)&0.07(0.33)&1.00(0.00)&
			2.99(0.08)&0.29(0.99)&0.99(0.08)\\ \hline
			\multirow{2}{*}{\textsc{I-SIS}}&500 & 3.00(0.00)&0.04(0.20)&1.00(0.00)&
			3.00(0.00)&0.21(0.91)&1.00(0.00)\\
			&1000&3.00(0.00)&0.05(0.27)&1.00(0.00)&
			3.00(0.00)&0.31(1.18)&1.00(0.00)\\ \hline
			\multirow{2}{*}{\textsc{C-FS}}&500 &3.00(0.00)&2.11(1.87)&1.00(0.00)&
			3.00(0.00)&2.02(1.73)&1.00(0.00)\\
			&1000&3.00(0.00)&4.28(2.90)&1.00(0.00)& 3.00(0.00)&4.15(2.91)&1.00(0.00)\\ \hline
			\multirow{2}{*}{\textsc{HOLP}}&500 & 2.95(0.26)&0.02(0.12)&0.95(0.21)&
			1.92(0.85)&0.12(0.33)&0.31(0.46)\\
			&1000&2.95(0.30)&0.01(0.04)&0.98(0.15)& 1.52(0.84)&0.04(0.19)&0.23(0.42)\\ \hline
			\multirow{2}{*}{\textsc{FARMS}}&500 & 3.00(0.00)&1.32(1.24)&1.00(0.00)&
			2.99(0.10)&1.47(1.34)&0.99(0.10)\\
			&1000&3.00(0.00)&1.45(1.28)&1.00(0.00)& 2.99(0.10)&1.67(1.39)&0.99(0.10)\\
			\hline
		\end{tabular}
	}
\end{table}

Tables \ref{tab1}-\ref{tab4} present the model selection results corresponding to Example \ref{lin}. In nearly all scenarios, the proposed T-DF algorithm consistently outperforms the other screening methods. Each method exhibits its highest selection efficiency in terms of TP, FP, and CR when the covariates are uncorrelated.

As shown in Table \ref{tab1}, at $\rho=0.5$, LASSO and FBIC exhibit slight underfitting, potentially leading to the selection of models that may not entirely include the true predictors. Meanwhile, HOLP's selected model shows a sharp decline in coverage probability, which is particularly noticeable at $p_n=1,000$. When $\rho=0.8$, LASSO, FBIC, SCAD, HOLP, and FARMS encounter challenges in identifying the true variables,  resulting in extremely low coverage ratios. I-SIS selects an excessive number of unimportant variables, leading to a coverage ratio lower than that achieved by our T-DF algorithm. If $\rho$ increases, the assumptions underlying marginal correlation and the irrepresentable condition may become invalid.

In Table \ref{tab2}, unlike in Table \ref{tab1}, all methods exhibit improved performance under the block compound symmetry correlation structure. The proposed T-DF algorithm achieves a coverage rate approaching 90\% even when $\rho=0.8$ and $p_n=1,000$. I-SIS and C-FS identify most of the signals at the cost of including some redundant covariates. Unfortunately, the screening efficiencies of LASSO, FBIC, SCAD, HOLP, and FARMS are adversely affected by variable correlation.

When the sample size increases to $n=400$, Tables \ref{tab3} and \ref{tab4} exhibit a similar trend and higher selection efficiency compared with that in Tables \ref{tab1} and \ref{tab2}, respectively. In Table \ref{tab4}, for $\rho=0.8$, TP increases to the number of true signals with minimal FP, indicating that the selected model almost certainly contains the true model. In the cases with autoregressive correlated covariates, the coverage rate of SCAD rises to greater than 20\%, and I-SIS and C-FS almost identify all significant covariates in 600 repetitions. FBIC's recognition efficiency  for the true model is considerably improved by up to 70\%. Meanwhile, the coverage ratio of FARMS is around 50\% when $p_n=500$ and is less than  a quarter when $p_n=1,000$. However, neither LASSO nor HOLP, with their extremely low coverage ratios is efficient when $p_n=1,000$ .

On the basis of these findings, we conclude that our T-DF algorithm can perform robust model selection under high-dimensional linear models when the covariates are highly correlated. The T-DF algorithm achieves screening consistency by correctly identifying important variables and excluding irrelevant ones with a probability approaching one.

\begin{table}[htb]
	\centering
%	\scriptsize
	\caption{TP, FP, and CR ($\%$) over 600 repetitions and standard deviations for logistic regression in Scenario (I) with $n=200$}
	\label{tab5}
	\vspace{-0.2cm}
	\resizebox{1\textwidth}{!}{
		\begin{tabular}{cc|ccccccccc}
			\hline
			&&TP&FP&CR&TP&FP&CR&TP&FP&CR\\ \hline
			Approach &$p_n$  &\multicolumn{3}{c}{$\rho=0$}& \multicolumn{3}{c}{$\rho=0.5$}& \multicolumn{3}{c}{$\rho=0.8$}\\ \hline
				\multirow{2}{*}{\textsc{T-DF}}&500 & 2.92(0.27)&1.69(0.66)&0.92(0.27)
			&2.54(0.83)&1.97(0.97)&0.73(0.44)&
			2.01(0.93)&2.19(1.00)&0.41(0.49)\\
			&1000&2.87(0.36)&1.60(0.63)&0.88(0.33)& 2.27(1.01)&2.85(1.15)&0.61(0.49)&
			1.84(0.92)&2.94(1.03)&0.30(0.46)\\ \hline
			\multirow{2}{*}{\textsc{LASSO}}&500 & 2.47(0.76)&0.18(0.46)&0.62(0.49)&
			0.66(0.64)&0.04(0.22)&0.00(0.00)&
			0.85(0.61)&0.10(0.34)&0.00(0.00)\\
			&1000&2.23(0.85)&0.13(0.38)&0.46(0.50)
			&0.48(0.62)&0.02(0.16)&0.00(0.00)&
			0.75(0.63)&0.07(0.27)&0.00(0.00)\\ \hline
			\multirow{2}{*}{\textsc{SCAD}}&500 & 2.66(0.54)&0.32(0.65)&0.69(0.46)& 0.96(0.64)&0.19(0.44)&0.01(0.04)&
			1.11(0.57)&0.23(0.49)&0.00(0.00)\\
			&1000&2.49(0.63)&0.21(0.47)&0.56(0.50)
			&0.82(0.64)&0.18(0.44)&0.00(0.00)&
			1.00(0.63)&0.22(0.50)&0.00(0.00)\\ \hline
			\multirow{2}{*}{\textsc{I-SIS}}&500 & 2.84(0.57)&5.52(1.69)&0.91(0.29)&
			0.53(0.96)&0.79(2.08)&0.10(0.30)&
			0.56(0.74)&0.42(1.45)&0.03(0.17)\\
			&1000&2.86(0.51)&5.88(1.09)&0.90(0.30)
			&0.77(1.18)&1.57(2.74)&0.18(0.38)&
			0.62(0.85)&1.25(2.57)&0.06(0.24)\\ \hline
			\multirow{2}{*}{\textsc{C-FS}}&500 & 2.79(0.47)&6.58(3.60)&0.82(0.39)&
			2.06(1.05)&5.90(3.36)&0.51(0.50)&
			1.16(0.62)&3.58(2.26)&0.06(0.24)\\
			&1000&2.73(0.51)&18.22(9.00)&0.76(0.43)& 1.91(1.08)&14.97(7.58)&0.45(0.50)&
			1.09(0.58)&8.11(4.45)&0.05(0.21)\\ \hline
			\multirow{2}{*}{\textsc{FARMS}}&500 & 2.87(0.42)&1.99(1.39)&0.90(0.31)&
			1.58(0.62)&3.87(1.82)&0.04(0.20)&
			1.28(0.59)&3.23(1.83)&0.00(0.00)\\
			&1000&2.84(0.40)&2.76(1.74)&0.85(0.35)& 1.44(0.64)&5.18(2.13)&0.02(0.12)&
			1.37(0.58)&3.80(1.98)&0.00(0.00)\\
			\hline
		\end{tabular}
}
\end{table}

\begin{table}[htb]
	\centering
%	\scriptsize
	\caption{TP, FP, and CR ($\%$) over 600 repetitions and standard deviations for logistic regression in Scenario (II) with $n=200$}
	\label{tab6}
	\vspace{-0.2cm}
\resizebox{0.8\textwidth}{!}{
		\begin{tabular}{cc|cccccc}
			\hline
			&&TP&FP&CR&TP&FP&CR\\ \hline
			Approach &$p_n$  & \multicolumn{3}{c}{$\rho=0.5$}& \multicolumn{3}{c}{$\rho=0.8$}\\ \hline
			\multirow{2}{*}{\textsc{T-DF}}&500 & 2.74(0.57)&2.09(0.81)&0.80(0.40)
			&2.19(0.79)&2.77(0.94)&0.40(0.49)\\
			&1000&2.65(0.58)&2.35(0.83)&0.71(0.45)& 2.04(0.79)&2.97(0.92)&0.30(0.46)\\ \hline
			\multirow{2}{*}{\textsc{LASSO}}&500 & 1.29(0.71)&0.06(0.24)&0.03(0.18)&
			1.13(0.56)&0.23(0.67)&0.00(0.00)\\
			&1000&1.18(0.73)&0.06(0.23)&0.03(0.17)
			&1.03(0.62)&0.25(0.67)&0.00(0.00)\\ \hline
			\multirow{2}{*}{\textsc{SCAD}}&500 & 1.54(0.69)&0.24(0.66)&0.08(0.27)& 1.17(0.50)&0.53(0.96)&0.00(0.00)\\
			&1000&1.42(0.67)&0.17(0.50)&0.05(0.22)
			&1.14(0.52)&0.56(1.04)&0.00(0.00)\\ \hline
			\multirow{2}{*}{\textsc{I-SIS}}&500 & 1.49(1.13)&1.64(2.65)&0.27(0.44)&
			0.87(0.59)&0.66(1.75)&0.01(0.11)\\
			&1000&1.33(1.12)&1.70(2.77)&0.22(0.41)
			&0.75(0.64)&1.04(2.42)&0.01(0.11)\\ \hline
			\multirow{2}{*}{\textsc{C-FS}}&500 & 2.21(0.79)&4.91(3.66)&0.43(0.50)&
			1.34(0.63)&4.17(3.61)&0.06(0.23)\\
			&1000&2.14(0.79)&14.85(9.32)&0.39(0.49)& 1.28(0.58)&12.98(9.78)&0.03(0.17)\\ \hline
			\multirow{2}{*}{\textsc{FARMS}}&500 & 2.36(0.63)&2.90(1.64)&0.44(0.50)&
			1.51(0.54)&3.31(1.74)&0.00(0.00)\\
			&1000&2.29(0.61)&3.88(1.93)&0.38(0.48)& 1.53(0.54)&4.40(2.02)&0.00(0.00)\\
			\hline
		\end{tabular}
}
\end{table}

\begin{table}[htb]
	\centering
%	\scriptsize
	\caption{TP, FP, and CR ($\%$) over 600 repetitions and standard deviations for logistic regression in Scenario (I) with $n=400$}
	\label{tab7}
	\vspace{-0.2cm}
\resizebox{1\textwidth}{!}{
		\begin{tabular}{cc|ccccccccc}
			\hline
			&&TP&FP&CR&TP&FP&CR&TP&FP&CR\\ \hline
			Approach &$p_n$  &\multicolumn{3}{c}{$\rho=0$}& \multicolumn{3}{c}{$\rho=0.5$}& \multicolumn{3}{c}{$\rho=0.8$}\\ \hline
				\multirow{2}{*}{\textsc{T-DF}}&500 & 3.00(0.00)&1.73(0.55)&1.00(0.00)
			&3.00(0.00)&1.23(0.50)&1.00(0.00)&
			2.80(0.51)&1.55(0.72)&0.85(0.36)\\
			&1000&3.00(0.00)&1.26(0.46)&1.00(0.00)& 3.00(0.00)&1.94(0.54)&1.00(0.00)&
			2.80(0.50)&2.57(0.77)&0.85(0.36)\\ \hline
			\multirow{2}{*}{\textsc{LASSO}}&500 & 3.00(0.00)&0.25(0.51)&1.00(0.00)&
			1.56(0.66)&0.10(0.35)&0.07(0.25)&
			1.52(0.51)&0.13(0.37)&0.00(0.00)\\
			&1000&2.99(0.10)&0.24(0.53)&0.99(0.10)
			&1.45(0.61)&0.07(0.28)&0.03(0.16)&
			1.46(0.52)&0.11(0.34)&0.00(0.00)\\ \hline
			\multirow{2}{*}{\textsc{SCAD}}&500 & 3.00(0.00)&0.49(0.71)&1.00(0.00)& 1.93(0.72)&0.84(1.50)&0.23(0.42)&
			1.61(0.49)&0.26(0.53)&0.00(0.00)\\
			&1000&3.00(0.00)&0.48(0.75)&1.00(0.00)
			&1.75(0.62)&0.49(1.30)&0.10(0.29)&
			1.56(0.51)&0.22(0.48)&0.00(0.00)\\ \hline
			\multirow{2}{*}{\textsc{I-SIS}}&500 & 3.00(0.07)&0.71(1.54)&1.00(0.00)&
			1.43(1.01)&0.81(2.31)&0.18(0.39)&
			1.23(0.68)&0.12(0.36)&0.00(0.00)\\
			&1000&2.99(0.08)&1.77(3.85)&0.99(0.08)
			&1.24(1.02)&1.58(4.13)&0.14(0.35)&
			1.21(0.71)&0.21(1.21)&0.01(0.10)\\ \hline
			\multirow{2}{*}{\textsc{C-FS}}&500 & 3.00(0.00)&5.36(2.85)&1.00(0.00)&
			2.96(0.27)&4.93(2.72)&0.98(0.15)&
			1.71(0.93)&3.42(2.13)&0.33(0.47)\\
			&1000&3.00(0.00)&12.01(5.09)&1.00(0.00)& 2.96(0.27)&10.81(4.78)&0.98(0.15)&
			1.71(0.93)&6.87(3.42)&0.33(0.47)\\ \hline
			\multirow{2}{*}{\textsc{FARMS}}&500 & 2.96(0.28)&1.38(1.19)&0.98(0.14)&
			2.58(0.54)&3.76(2.01)&0.60(0.49)&
			1.56(0.52)&3.31(1.93)&0.00(0.00)\\
			&1000&3.00(0.00)&1.79(1.35)&1.00(0.00)& 2.41(0.58)&4.86(2.23)&0.45(0.50)&
			1.63(0.49)&3.98(2.10)&0.00(0.00)\\
			\hline
		\end{tabular}
}
\end{table}

\begin{table}[htb]
	\centering
%	\scriptsize
	\caption{TP, FP, and CR ($\%$) over 600 repetitions and standard deviations for logistic regression in Scenario (II) with $n=400$}
	\label{tab8}
	\vspace{-0.2cm}
\resizebox{0.8\textwidth}{!}{
		\begin{tabular}{cc|cccccc}
			\hline
			&&TP&FP&CR&TP&FP&CR\\ \hline
			Approach &$p_n$  & \multicolumn{3}{c}{$\rho=0.5$}& \multicolumn{3}{c}{$\rho=0.8$}\\ \hline
		\multirow{2}{*}{\textsc{T-DF}}&500 & 2.99(0.09)&1.73(0.53)&0.99(0.09)
		&2.89(0.34)&2.73(0.70)&0.89(0.31)\\
		&1000&2.99(0.10)&2.05(0.49)&0.99(0.10)& 2.82(0.41)&2.95(0.75)&0.82(0.38)\\ \hline
		\multirow{2}{*}{\textsc{LASSO}}&500 & 2.49(0.66)&0.21(0.52)&0.59(0.49)&
		1.66(0.47)&0.08(0.31)&0.00(0.00)\\
		&1000&2.38(0.70)&0.23(0.55)&0.51(0.50)
		&1.62(0.49)&0.11(0.42)&0.00(0.00)\\ \hline
		\multirow{2}{*}{\textsc{SCAD}}&500 & 2.74(0.47)&0.71(1.00)&0.75(0.43)& 1.21(0.42)&0.29(0.65)&0.01(0.08)\\
		&1000&2.60(0.68)&0.73(1.13)&0.65(0.48)
		&1.20(0.42)&0.37(0.80)&0.01(0.08)\\ \hline
		\multirow{2}{*}{\textsc{I-SIS}}&500 & 2.68(0.57)&0.84(1.14)&0.73(0.45)&
		1.17(0.46)&0.25(0.61)&0.02(0.15)\\
		&1000&2.58(0.64)&1.31(2.58)&0.66(0.48)
		&1.13(0.44)&0.30(0.85)&0.01(0.11)\\ \hline
		\multirow{2}{*}{\textsc{C-FS}}&500 & 2.94(0.26)&4.57(3.02)&0.95(0.22)&
		2.03(0.82)&4.29(3.06)&0.35(0.48)\\
		&1000&2.92(0.30)&10.69(5.71)&0.93(0.26)& 2.03(0.80)&10.17(6.14)&0.34(0.47)\\ \hline
		\multirow{2}{*}{\textsc{FARMS}}&500 & 2.98(0.15)&2.38(1.60)&0.98(0.15)&
		1.86(0.43)&3.54(1.88)&0.03(0.18)\\
		&1000&2.95(0.22)&3.22(1.85)&0.95(0.22)&1.83(0.42)&4.47(2.21)&0.02(0.13)\\
			\hline
		\end{tabular}
}
\end{table}

\begin{Example}{\rm  \label{los}%(\textit{high dimensional linear regression})
	In the logistic regression model, we define the probability of $Y_i$ as 1 given $\bx_i$ as $\pi_{i}$.  This relationship is captured by the equation:
	\begin{equation*}
		P(Y_i=1| \bx_i)=\pi_{i},\ \ \log(\pi_{i}/(1-\pi_{i}))=\bx^\top_i\bbeta^*
	\end{equation*}
for  $i=1,\dots, n$. Here,  $\bx_i$s are $i.i.d$  from $N(\b0,\bSigma)$, with $\bSigma$ having the same structure as that described in Example
\ref{lin}. The true regression coefficients over the $n$ observations are set as $\bbeta^*=(1,-1,0.8,\textbf{0}_{p_n -3})^\top$.
	}
\end{Example}

Tables \ref{tab5}-\ref{tab8}  display the screening performance of Example \ref{los}. In contrast to linear regression, logistic regression introduces heteroscedastic  errors, which complicate the identification of important predictors. When the sample size increases, all methods show improved screening efficiency.
Overall, the proposed T-DF algorithm consistently outperforms the other screening approaches.
%The proposed T-DF algorithm consistently outperforms others across all scenarios. 
At a sample size of $400$,  it achieves consistent screening results with an overwhelming probability at  $\rho=0.5$ and maintains a relatively robust performance even when correlation $\rho$ increases to $0.8$.

For the uncorrelated features, SCAD, LASSO, I-SIS, C-FS, and FARMS exhibit consistent abilities in variable selection at a sample size of $n=400$. However, this capability diminishes gradually with increasing correlation between covariates to the extent that when correlation $\rho$ increases to $0.8$, their CR values approach zero or become very low. For instance, in Scenario (I) with $p_n=500$, I-SIS achieves 18\% CR at  $\rho=0.5$, which diminishes to 0 at $\rho=0.8$. Similarly, in Scenario (II), I-SIS exhibits 73\% CR at $\rho=0.5$ with $p_n=500$, which decreases to 2\% at $\rho=0.8$. By comparison, the performance of C-FS in identifying true features is substantially inferior to that of our T-DF algorithm because it tends to overfit by selecting unimportant variables.
The performance of FARMS is also influenced severely by the strong correlation among features in Scenario (I). The selection efficiency of FARMS is comparable to that of the proposed T-DF algorithm in Scenario (II) at $\rho=0.5$. However, its screening ability vanishes when high correlation exists among the predictors.

In summary,  the proposed T-DF algorithm demonstrates superior model selection efficiency in high-dimensional logistic regression models, particularly  with highly correlated predictors, compared with the other approaches. Extensive numerical studies confirm its screening consistency even with increasing sample size.

\section{Real data applications}\label{sec:real}
In this section, we present two real data analysis examples to demonstrate our methodology.
\subsection{\textit{Arabidopsis thaliana} gene data}\label{sec:real1}
\begin{figure}[hb]
%	\centering
\begin{subfigure}[t]{0.5\textwidth}
		\centering
	\includegraphics[height=5.5cm,width=8cm]{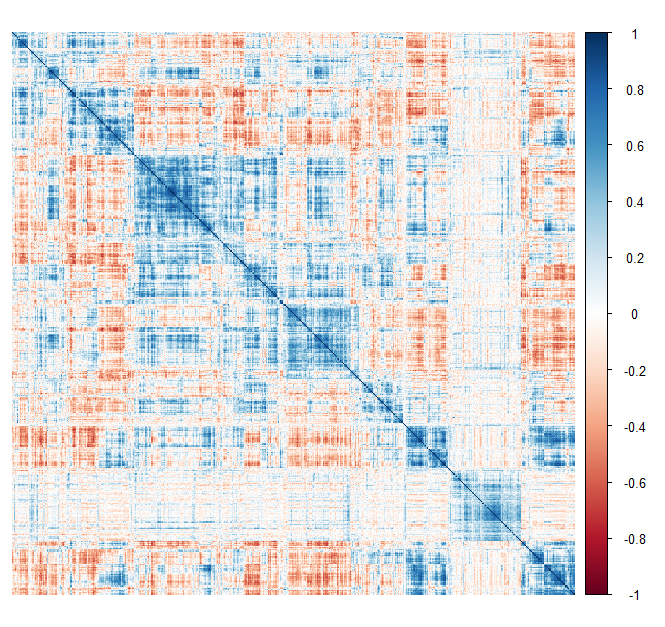}
		\vspace{-0.6cm}
	\subcaption*{(a) Correlation heat map}
\end{subfigure}
\begin{subfigure}[t]{0.5\textwidth}
		\centering
\includegraphics[height=5.5cm,width=8cm]{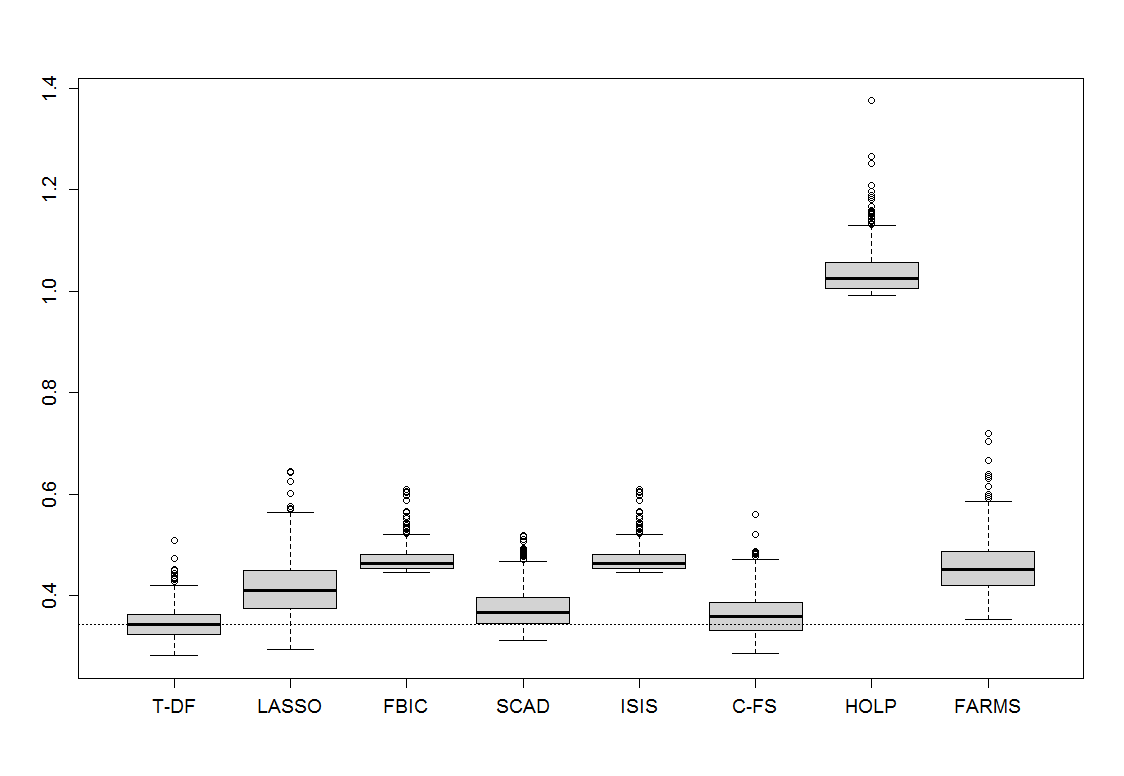}
	\vspace{-0.6cm}
	\subcaption*{(b) Box-plots of  prediction errors }
\end{subfigure}
%	\includegraphics[height=6cm,width=7cm]{833.png}
%		\includegraphics[height=6cm,width=8.5cm]{at_error.png}
%	%		\vspace{-0.6cm}
		\vspace{-0.6cm}
	\caption{\label{at_cor}Correlation heat map and prediction errors for \textit{Arabidopsis thaliana} gene data}
\end{figure}

First, we employ  \textit{Arabidopsis thaliana} gene data to show the screening performance of our T-DF algorithm. This dataset is accessible in the supplementary materials of \cite{Wille04} (\url{https://www.ncbi.nlm.nih.gov/pmc/articles/PMC545783/})
 and 
comprises $n=118$ samples, each containing values of $p_n=834$ genes across 58 distinct pathways. 
Biologically, some key compounds, such as carotenoids, chlorophylls, and gibberellins, use geranylgeranyl diphosphate (GGPP) as a precursor in Arabidopsis. \cite{C2015} indicated that GGPP generation primarily relies on the gene GGPPS11.  In this study,  our aim is to investigate how the expression level of GGPPS11 is influenced by the 833 other genes. We present the correlation heatmap of the 833 genes in Figure \ref{at_cor}(a). The heatmap displays a nearly block-diagonal structure, indicating that a strong correlation exists among the expression levels of most genes.

We begin by standardizing the gene dataset and fitting it with the linear model:
\begin{equation}
	Y_i=\sum_{j=1}^{p_n}x_{ij}\beta^*_j+ \epsilon_i,\qquad i=1,\dots,n, \label{real}
\end{equation}
where $Y_i$ represents the expression values of gene GGPPS11 and $x_{ij}$ denotes the expression value of the $j$th remaining gene in the $i$th sample. 
Next, we apply feature selection techniques (outlined in Section \ref{sec5}) to the dataset, and the detailed results are presented in Table \ref{tab9}. To further evaluate the performance of these methods, we implement a cross-validation approach on the dataset and compare the prediction errors.

For this purpose, we randomly split the entire dataset equally into two subsamples labeled as $\mathcal{A}_1$ and $\mathcal{A}_2$. Given submodel $\hat{\mathcal{S}}$, we compute the least squares estimator $\hat\bbeta_{\hat{\mathcal{S}}}^{(k)}$ by using these subsamples. The prediction error is then given by
$$\frac{1}{n}\sum_{k=1}^2\sum_{i\in \mathcal{A}_k} \bigl\{Y_i-\bx_{i,\hat{\mathcal{S}}}^\top\hat\bbeta_{\hat{\mathcal{S}}}^{(2)}\bigr\}^2.$$

\begin{table}	
	\centering
	\caption{\label{tab9} Selected model size of eight considered methods }
		\vspace{-0.2cm}
	\resizebox{0.9\textwidth}{!}{
			\begin{tabular}{ccccccccc}
				\hline
				Approach &  T-DF & LASSO & 	FBIC & SCAD &  I-SIS &C-FS &	HOLP-EBIC & FARMS \\ \hline 
				Model size  & $5$  & $12$  & $1$ & $4$  & $1$   & $9$   & $1$ & $7$ \\
					\hline
		\end{tabular}
	}
\end{table}

We repeat this procedure 600 times and visualize the prediction errors via a boxplot (Figure \ref{at_cor}(b)). Remarkably, our T-DF algorithm consistently favors a model of moderate size and exhibits a minimal prediction error. 
The LASSO method tends to favor a large model, leading to a  high prediction error. This observation indicates LASSO's limitation in discerning pivotal features,  possibly incorporating extraneous covariates.  
Meanwhile, FBIC, SCAD, and I-SIS manifest a pronounced prediction deviation, leading to a suboptimal fit of the genuine model.   
By contrast, C-FS and FARMS algorithms are inclined toward large models, which can introduce some instability. Their deflected prediction errors suggest the inclusion of redundant covariates. 
Additionally, the HOLP methodology exhibits substantial prediction errors for minor genes, signifying the omission of crucial variables and consequent underfitting.

In conclusion, our T-DF algorithm performs efficiently and robustly in identifying true genes that may regulate the expression of the gene GGPPS11.

\subsection{Breast cancer data}
Next, we demonstrate the proposed T-DF algorithm through the analysis of breast cancer data. Breast cancer is the second most prevalent cancer globally and poses a critical threat to human health, particularly among women.  Accurately predicting tumor metastasis is crucial for effective breast cancer treatment. Here, we use the gene expression data initially examined by \cite{Van02} to explore the influence of gene probes on breast cancer tumor metastasis.

The data comprise $n=97$ patients aged 55 or younger with lymph node-negative breast cancer. Among them, 46 patients developed distant metastases within five years (coded as one), and 51 remained metastasis-free for at least five years (coded as zero).

\begin{figure}[hb]
	%	\centering
	\begin{subfigure}[t]{0.5\textwidth}
		\centering
		\includegraphics[height=5.5cm,width=8cm]{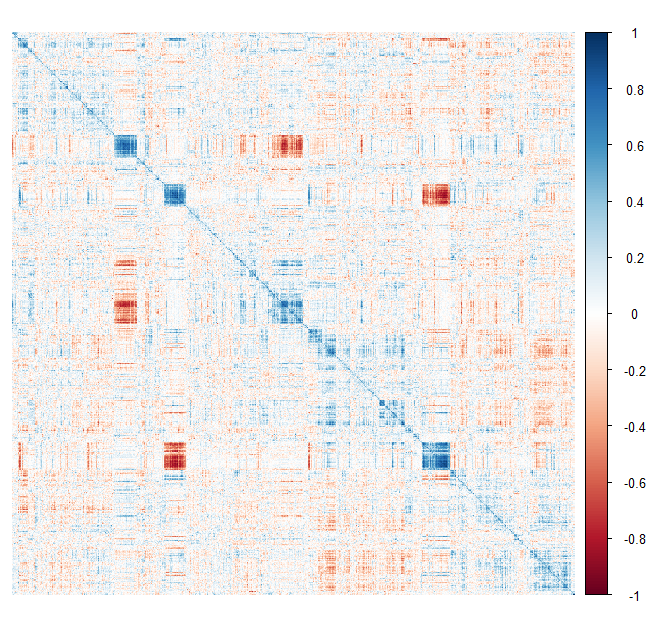}
		\vspace{-0.6cm}
		\subcaption*{(a) Correlation heat map}
	\end{subfigure}
	\begin{subfigure}[t]{0.5\textwidth}
		\centering
		\includegraphics[height=5.5cm,width=8cm]{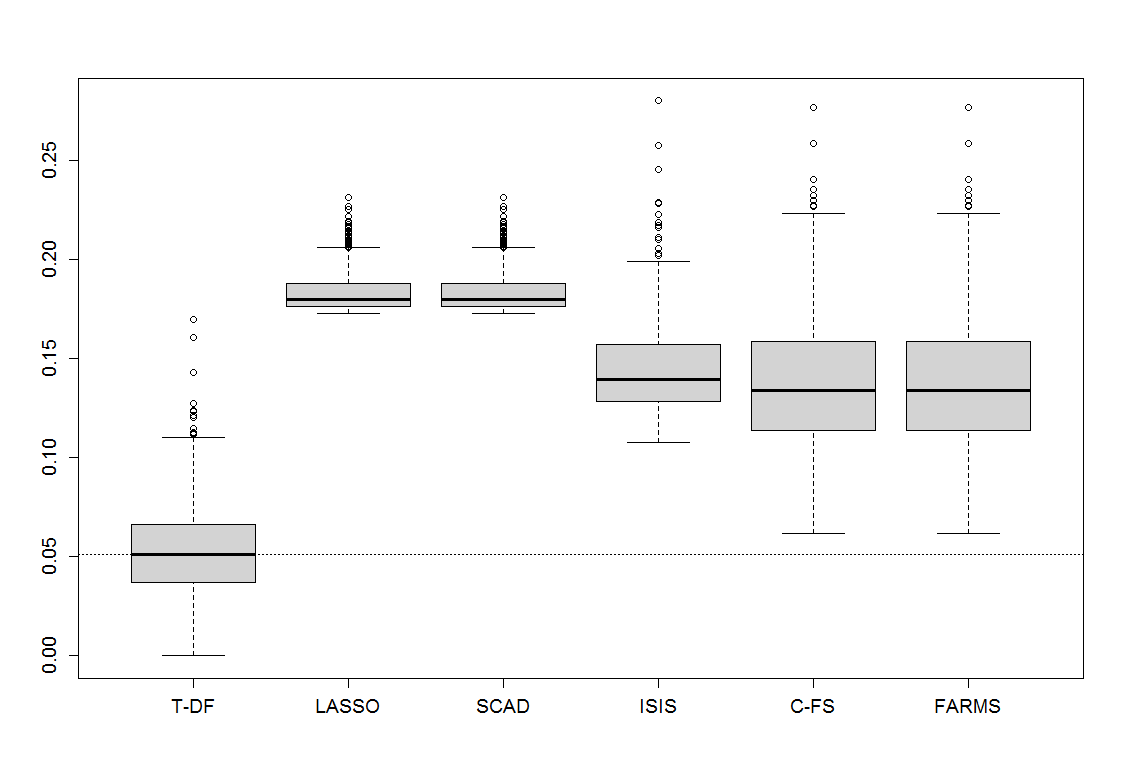}
		\vspace{-0.6cm}
		\subcaption*{(b) Box-plots of  prediction errors }
	\end{subfigure}
	%	\includegraphics[height=6cm,width=7cm]{833.png}
	%		\includegraphics[height=6cm,width=8.5cm]{at_error.png}
	%	%		\vspace{-0.6cm}
	\vspace{-0.6cm}
	\caption{\label{final_error}Correlation heat map and prediction errors for Breast cancer data}
\end{figure}

We analyze the expression levels of  $p_n=24,188$ gene probes without any missing values. Focusing on the top 2,000 genes with the highest variance, we generate a heatmap illustrating their estimated correlation matrix (Figure \ref{final_error}(a)). Notably, strong correlations are observed among these genes, underscoring the need for our decorrelation method.

\begin{table}[htb]
	\centering
	\caption{\label{tab11}Selected model size of six considered methods  }
	\vspace{-0.2cm}
	\resizebox{0.7\textwidth}{!}{
		\begin{tabular}{ccccccc}
			\hline
			Approach & T-DF & 	LASSO & SCAD &ISIS &C-FS &FARMS\\ \hline
			Model size & $8$  & $1$ & $1$ & $5$ & $22$ &$6$\\
			\hline
		\end{tabular}
	}
\end{table}

With Model (\ref{S2.1}), we conduct logistic regression to predict binary outcomes by using a large number of highly correlated covariates. The logistic regression model is expressed as
\begin{equation}
	\log\frac{\pi_{i}}{1-\pi_{i}}=\sum_{j=1}^{p_n}x_{ij}\beta^*_j, \qquad i=1,\dots,n, \label{log_mod}
\end{equation}  
where $\pi_{i}$ represents the marginal mean of $Y_i$ and $x_{ij}$ is the expression value of the $j$th gene probe for the $i$th patient. 
We analyse the binary outcomes' data by using the methods mentioned in Section \ref{sec5}. We present the selected model sizes for each approach in Table \ref{tab11}.

The breast cancer data are randomly divided into two parts denoted as $\mathcal{B}_1$ and $\mathcal{B}_{2}$. Given submodel $\hat{\mathcal{S}}$, we compute the maximum likelihood estimator $\hat\bbeta_{\hat{\mathcal{S}}}^{(k)}$ by using sub-sample $\mathcal{B}_{k}$ $(k=1,2)$. In logistic regression, the prediction error is calculated as
$$n^{-1}\sum_{k=1}^2\sum_{i\in \mathcal{B}_k}\bigl\{Y_i-\text{logit}(
\bx_{i,\hat{\mathcal{S}}}^\top\hat\bbeta_{\hat{\mathcal{S}}}^{(k)})
\bigr\}^2, \ \ \text{logit}(t)=\{1+\exp(-t)\}^{-1}.$$
We repeat the process 600 times and compare the prediction errors in Figure \ref{final_error}(b).
 
Similar to the results in Table \ref{tab9}, the proposed T-DF algorithm identifies a moderate-size model and has the smallest prediction error among the compared methods.  LASSO and SCAD select models that are too small, resulting in considerable prediction deviations, indicating their limitations in recognizing important features. I-SIS and FARMS determine a model size that is comparable to that of the T-DF algorithm. However, their prediction errors suggest their insufficient capability to accurately include numerous true features and exclude redundant gene probes. Although the prediction error of C-FS approaches that of the T-DF algorithm, it selects too many false features, leading to severe overfitting.

\section{Discussion}

We introduce a novel method to transform complex Model \eqref{b11} into linear Model \eqref{S2.1} and establish the DF regression method to identify the important variables. The value of this work lies in the fact that the proposed approaches use the least squared method to deal with  complex data and consider the strong correlation within features. Compared with existing sure screening methods and penalized model selection techniques, our methods elinimate the process of modeling conditional variance  with little model structure and show robust performance against correlation.

Post-selection inference plays a crucial role in high-dimensional inference. It involves a two-step process.  The first step is to employ  variable selection methods to reduce the dimension of feature spaces, and the second step focuses on constructing test statistics within these reduced dimensions. Recent research has concentrated on utilizing classical penalized and screening methods in the first step, with emphasis on constructing test statistics thereafter. However, as mentioned earlier, classical penalized and screening methods often require modeling conditional variance and are sensitive to predictor correlation.

Our methods offer a solution for implementing the first step in post-selection inference in complex data analyses and settings with strong correlations. Furthermore, the screening consistency and the upper bound of the selected model size  provide theoretical guarantees for inference in the second step. Such advancements can facilitate the widespread adoption of post-selection inference in the analysis of complex real-world data.

\spacingset{1.1}

%---------------------------------------------------------

\end{document}